\documentclass[journal,draftclsnofoot,onecolumn,12pt,twoside]{IEEEtran}
\normalsize
\usepackage[T1]{fontenc} 
\usepackage{cite}
\usepackage{amsmath,amssymb,amsfonts}
\usepackage{mathabx}
\usepackage{fdsymbol}
\usepackage{xfrac}
\usepackage{algorithmic}
\usepackage[font=small,labelfont=bf,skip=0pt]{caption}
\usepackage{subcaption}
\usepackage{algorithm}
\usepackage{enumitem}
\usepackage{tabularx}
\usepackage{graphicx}
\usepackage{xcolor}
\usepackage{amsthm}
\theoremstyle{remark}
\newtheorem{remark}{Remark}
\newtheorem{thm}{Theorem}

\renewcommand{\theenumi}{\Alph{enumi}}
\newtheorem{prop}{Proposition}

\ifCLASSINFOpdf
\else
 
\fi

\usepackage{amsmath}

\begin{document}
\title{Multi-Transmitter Coded Caching with Secure Delivery over Linear Networks\\
--- Extended Version ---}

\author{Mohammad Javad Sojdeh, \textit{Student Member, IEEE}, Mehdi Letafati, \textit{Student Member, IEEE}, Seyed Pooya Shariatpanahi, \textit{Member}, \textit{IEEE} and Babak Hossein Khalaj, \textit{Senior Member}, \textit{IEEE}% <-this % stops a space
\thanks{Some preliminary results were presented at the 23rd IEEE International Workshop on Signal Processing Advances in Wireless Communications (SPAWC 2022), Oulu, Finland, July 4-6, 2022 \cite{spawc_2022}.}\\
\thanks{ M. J. Sojdeh and S. P. Shariatpanahi are with School of Electrical and Computer Engineering, College of Engineering, University of Tehran, 1439957131 Iran, e-mails: {\{sojdehjavad1376, p.shariatpanahi\}@ut.ac.ir}.}% <-this % stops a space
\thanks{M. Letafati and B. H. Khalaj are with the Department of Electrical Engineering, Sharif University of Technology, Tehran 1365-11155, Iran, e-mails: {\{mletafati@ee.sharif.edu; khalaj@sharif.edu\}}.}}

%\markboth{Journal of \LaTeX\ Class Files,~Vol.~14, No.~8, August~2015}%
%{Shell \MakeLowercase{\textit{et al.}}: Secure Multi-Antenna Coded Caching}
\maketitle
\vspace{-20mm}
\date{}
\begin{abstract}
  In this paper, we consider multiple cache-enabled end-users connected to multiple transmitters through a linear network. We also prevent a totally passive eavesdropper, who sniffs the packets in the delivery phase, from obtaining any information about the original files in cache-aided networks. Three different secure centralized multi-transmitter coded caching scenarios namely, secure multi-transmitter coded caching, secure multi-transmitter coded caching with reduced subpacketization, and secure multi-transmitter coded caching with reduced feedback, are considered and closed-form coding delay and secret shared key storage expressions are provided. As our security guarantee, we show that the delivery phase does not reveal any information to the eavesdropper using the mutual information metric. Moreover, we investigate the secure decentralized multi-transmitter coded caching scenario, in which there is no cooperation between the clients and transmitters during the cache content placement phase and study its performance compared to the centralized scheme. We analyze the system's performance in terms of \emph{Coding Delay} and guarantee the security of our presented schemes using the \emph{Mutual Information} metric. Numerical evaluations verify that security incurs a negligible cost in terms of memory usage when the number of files and users are scaled up, in both centralized and decentralized scenarios. Also, we numerically show that by increasing the number of files and users, the secure coding delay of centralized and decentralized schemes became \emph{asymptotically} equal.
\end{abstract}

\IEEEpeerreviewmaketitle

\section{Introduction}
\vspace{-1mm}
\IEEEPARstart{E}{mergence} of new data-intensive services such as wireless extended reality, virtual and augmented reality (XR/VR/AR) applications \cite{xr_6G} as the key $6$G drivers, wireless communication networks are under mounting pressure to achieve higher data rates and lower latency. In the following years, a variety range of wireless XR and Metaverse applications for social networking, gaming, education, and industry has emerged, pushing the capabilities of the underlying communication infrastructure. Utilizing novel multi-transmitter coded caching techniques is a promising approach to enable people to work, play and interact in a persist online $3$-D virtual environment with an immersive experience by generating an imaginary environment similar to the real world \cite{Extended_Reality_salehi,Non_Symmetric_salehi,Metaverse}. In particular, using caching and computing capabilities of XR mobile gadgets has effectively alleviated the traffic burden of wireless networks. Furthermore, significant bandwidth and delay-reduction gains have also been achieved \cite{xr_1,xr_2}.
\vspace{-6mm}
\subsection{Related Works}
\vspace{-1mm}
In addition to the conventional local caching gain, the seminal work of Maddah-Ali and Niesen \cite{Maddah_2014} reveals how coded caching can provide global caching gain by creating multi-casting opportunities using the coding technique. Their proposed scheme suggests storing non-identical file chunks among different end-users in the off-peak hours, and sending coded messages to serve multiple clients concurrently at the high-peak hours. By designing a careful pattern in the cache of end-users, and exploiting the coding techniques in the delivery phase, a significant global caching gain can be attained which is considerably more than the local caching gain achieved by the local cache memory of each user. Following this paradigm, decentralized coded caching scheme based on independent random content placement was introduced by Maddah-Ali and Niesen \cite{Maddah_2015-decentralized}. It was shown that, in the large file size regime, the multi-casting gain is almost the same as the centralized caching scenario for large networks.

Along another research line, the authors in \cite{Secure Delivery} propose the secure caching scheme with the security constraint of minimizing the information leakage to a totally passive eavesdropper during the delivery phase. They investigate cache memory vs. transmission rate trade-offs in centralized and decentralized scenarios. Their results indicate that security can be achieved at a negligible cost in terms of required storage, especially for large number of files and users. As well, they show their proposed achievable rate is within a constant multiplicative factor from the optimal rate achieved by the information-theoretic approach for almost all parameter values of interest.

In a follow-up paper, \cite{D2D Secure caching}, the authors study the fundamental limits of device-to-device (D2D) coded caching systems with security guarantees, where a user must not access files that are not requested. They use non-perfect secret sharing and one-time pad keying to provide security guarantees. In \cite{Secure-combinatorial method}, the authors propose a secure coded caching scheme to prevent wiretappers from obtaining any useful information about the files. They utilize a secure placement delivery array (SPDA) to characterize their scheme. The study in \cite{Colluding-Users} proposes a novel caching scheme to protect data security. When any subset of users collude and share all contents in their memories, they cannot achieve any information about the original files that they have not requested.

On a different line of research, plenty of works are proposed to enhance the performance of coded caching by using multiple transmitters/antennas \cite{Shariatpanahi_2016,Multi-antenna-coded-caching,Physical-Layer Schemes}. In cache-aided networks with multiple transmitters, the global caching gain, and the spatial multiplexing gain are aggregated with the help of the zero-forcing technique. In \cite{Feedback-bottleneck}, the authors show that multiplexing gain is optimal with the assumptions of uncoded prefetching, and one-shot data delivery.
\vspace{-4.8mm}
\subsection{Our Contributions}
\vspace{-2mm}
This paper investigates the fundamental security aspects of the \emph{multi-transmitter} coded caching problem in the presence of a \emph{totally passive eavesdropper}, for both \emph{centralized}, and \emph{decentralized} scenarios. In secure multi-transmitter caching problem, the multi-cast communication between the transmitters, and end-users happens over an insecure channel, and we use secret shared keys to encrypt each multi-cast message in the delivery phase. The fundamental cache memory vs. coding delay trade-off for the secure multi-transmitter caching problem is characterized. Our work generalizes the secure coded caching scenario \cite{Secure Delivery} to a multi-transmitter setup. Our paper is the first work in the context of security in cache-aided networks with multiple transmitters. Different from \cite{Shariatpanahi_2016}, we propose the secure centralized multi-transmitter coded caching scheme, and investigate the key storage requirement. In addition, we investigate the security integration into two state-of-the-art centralized multi-transmitter coded caching schemes \cite{Adding_transmitters}, and \cite{Feedback-bottleneck}. The motivation behind proposing the secure version of \cite{Adding_transmitters,Feedback-bottleneck} is that they benefit from having reduced feedback and subpacketization. Although the secure version of centralized algorithms \cite{Adding_transmitters,Feedback-bottleneck} require larger memory storage, as our security cost, compared with our first algorithm, they have the advantage of having reduced feedback, and subpacketization level, as two major concerns of cache-aided networks. Then, we compare the performance of these three secure centralized multi-transmitter proposed schemes in terms of coding delay and secure key storage size. Also, different from \cite{WSA2021}, a secure decentralized multi-transmitter caching scenario is presented, and analyzed. Moreover, we derive the closed-form expression of the secure key storage requirement for the generic case of secure multi-transmitter coded caching in both centralized, and decentralized settings. We analyze the system’s performance in terms of \emph{Coding Delay},  and guarantee the security of our presented schemes using the \emph{Mutual Information} metric for both centralized, and decentralized schemes. To satisfy our security constraint, we utilize uniformly distributed orthogonal keys, according to the one-time pad keying \cite{one_time_pad}, which are cached across the end-users for the security of multi-cast delivery. Our numerical results illustrate that by increasing the number of files and users, security can be achieved at no considerable cost in terms of memory usage for both centralized, and decentralized schemes. Interestingly, our numerical evaluations depict that the gap between the secure centralized, and decentralized schemes decreases by scaling the network. In other words, the secure coding delay of centralized and decentralized schemes become \emph{almost coincide} i.e., security from an eavesdropper can be achieved at \emph{almost negligible cost} for a large number of files and users \footnote{The optimality of the proposed schemes can be mathematically examined by presenting the converse bound of our achievable schemes, which will be investigated in our future works.}.

To summarize, the contributions of our work are as follows:
\begin{itemize}
	\item Three different secure centralized \emph{multi-transmitter} coded caching algorithms, are considered and closed-form coding delay, and secure key storage expressions are formulated. We compare these scenarios in terms of coding delay and key storage size.
	\item A secure decentralized \emph{multi-transmitter} caching scheme, in which each user fills its cache independent of other users during the placement phase, is presented and analyzed.
	\item We analyze the system's performance in terms of \emph{Coding Delay} and guarantee that security of our presented schemes using the \emph{Mutual Information} metric.
	\item We derive the closed-form expression of the key storage requirement for the generic case of secure multi-transmitter coded caching for both centralized and decentralized schemes. 
	\item Numerical results illustrate that by increasing the number of files and users, security can be achieved at no considerable cost in terms of memory, for both centralized and decentralized schemes. Moreover, we numerically show that the secure coding delay of the centralized and decentralized schemes are asymptotically equal.
\end{itemize}

Notations: We use $[{K}]$ to denote the set $\{{1},...,{K}\}$. Bold-face upper, and lowercase letters denote matrices, and vectors, respectively. Greek uppercase	letters are reserved for sets. For any matrix $\mathbf{A}$ (vector $\mathbf{a}$), ${\mathbf{A}^{T}}$ (vector ${\mathbf{a}^{T}}$) shows the transpose of $\mathbf{A}$ (vector $\mathbf{a}$). Also, $\mathbb{N}$, and ${{\mathbb{F}}_{q}}$ indicate the set of natural numbers, and a finite field with $q$ elements, respectively. $\mathbb{F}_{q}^{a\times b}$ represents the set of all ${a}\tiny{-}{by}\tiny{-}{b}$ matrices whose elements belong to ${{\mathbb{F}}_{q}}$. Moreover, $\oplus$ represents the addition in the corresponding finite field. Furthermore, ${\left( \begin{smallmatrix}
	K \\ 
	n \\ 
	\end{smallmatrix} \right)}$ denotes the n-choose-k (binomial) operator, and the
expression $\alpha |\beta$  denotes that integer $\alpha$ divides integer $\beta$. We generally reserve $I(X;Y)$ for the mutual information between $X$, and $Y$, and $H(X)$ for the entropy of $X$.

Paper Organization: The rest of the paper is organized as follows. Section II
presents the system model, and assumptions. Sections III, and IV introduce the centralized, and decentralized multi-transmitter coded caching algorithms, respectively. Section V presents numerical results. Finally, Section VI concludes the paper.
\vspace{-4.5mm}
\section{System Model and Assumptions}
This paper considers a content delivery network where $L$ transmitters are connected to $K$ cache-enabled end-users through a linear network \footnote{Definition of linear networks is provided in \cite{Shariatpanahi_2016}. In this type of networks, we assume that each node generates a random linear combination of data at its input ports to be transmitted on its output ports. The overall transmit and receive vectors of these networks are linearly related at each time slot.} \cite{Shariatpanahi_2016,Multi-antenna-coded-caching}, as shown in Figure 1. Each transmitter has access to the entire library, including $N$ files, where $N\ge K$. Also, each end-user in the network has a local cache memory with the size of $MF$ bits. The $F$-bit discrete files are represented by $m$-bit symbols over finite field GF($q$), where $m={\log_{2}}{q}$. Moreover, we assume that each file will include $\large{f\triangleq \sfrac{F}{m}}$ symbols (we suppose that $f$ is an integer).
\begin{figure}[t]
	\center
	\setlength\belowcaptionskip{0pt}
	\includegraphics[scale=0.25]{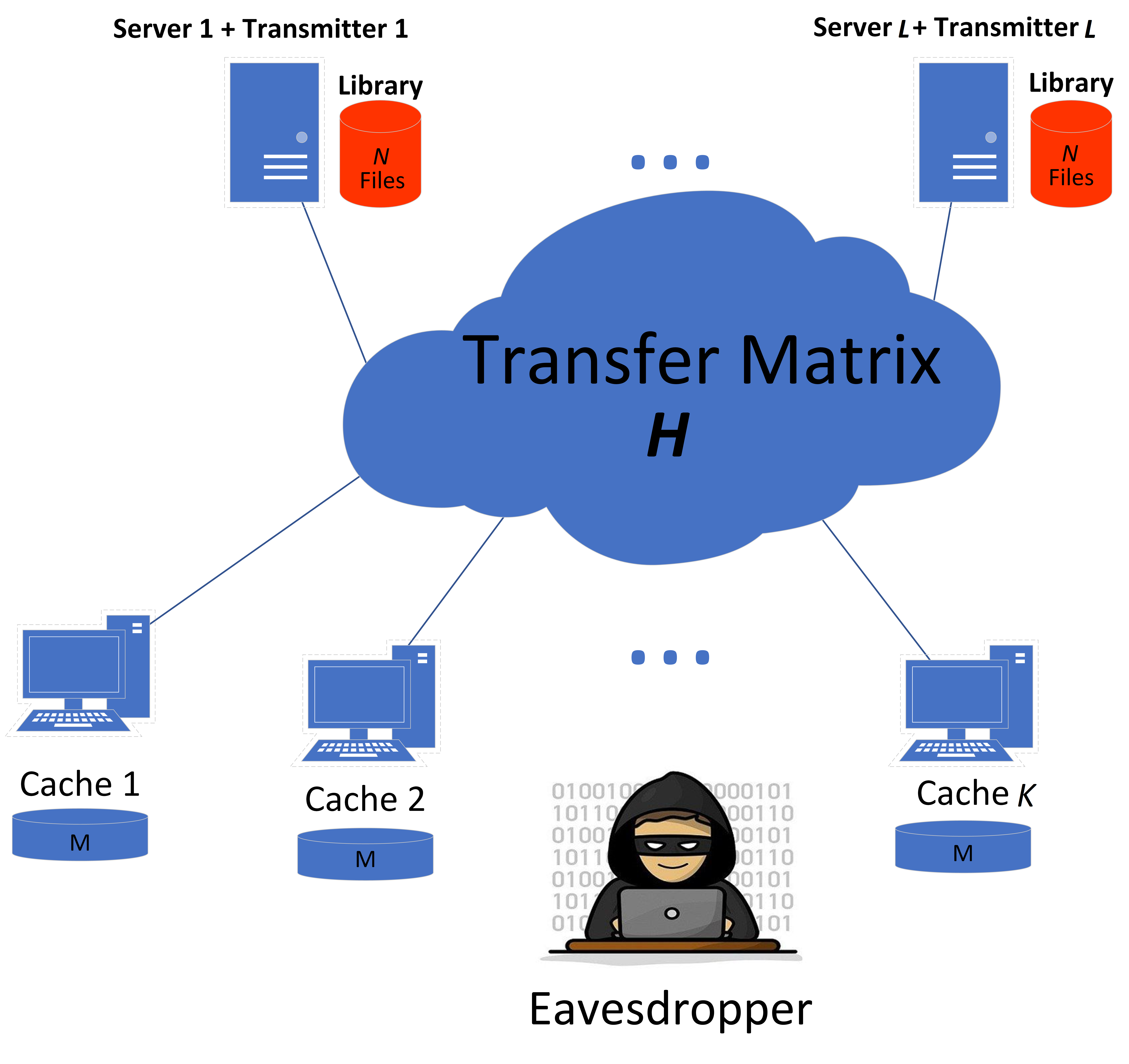}
	\caption{System Model: $L$ transmitters are connected to $K$ cache-enabled end-users via a linear network in the presence of a totally passive eavesdropper.
		\label{fig:System-Model}}
\end{figure}

We emphasize that the considered linear network has a generic structure, spanning from the wired to wireless systems. The random linear network coding approach, for instance, can be used to translate a wired coded caching problem into the linear network model described above \cite{linear_network_coding}. Furthermore, the compute-and-forward approach described in \cite{Compute-and-forward} can be applied for transforming the wireless caching scenario to the aforementioned finite field model.

The system experiences two distinct traffic states during its operation, namely off-peak, and high-peak hours. The prefetching step, also known as the \emph{Cache Content Placement} phase, can be of two types: \textbf{centralized} or \textbf{decentralized}, and takes place in the off-peak hours without any prior knowledge of the real requested files in the next phase. In the case of centralized placement, the transmitters fill the cache of each user with some parts of files. Each coordinating transmitter arranges the caches such that every subset of the cache memories shares a specific part of the content (i.e., there is a cooperation between each transmitter, and users in the placement phase) \cite{Maddah_2014}. In the case of decentralized placement, each user, independent of other users, caches any random combination of bits from each file without the coordination  from the transmitters \cite{Maddah_2015-decentralized}.

We use a slotted time model for the delivery phase, in which channel-uses in the network are represented by time slots $n=1,2,\ldots,{T}_{C}$. The second phase consists of ${T}_{C}$ time slots to deliver the requested files. At time slot $n$, during the second phase, the transmitted symbols, ${\mathbf{s}}(n)=[s_1(n) ,\ldots,s_L(n)]^T$ are sent from the $L$ transmitters, and end-users receive symbols ${\mathbf{y}_{r}}(n)=[{{y}_{{r}_1}}(n) ,\ldots,{{y}_{{r}_K}}(n)]^T$, and eavesdropper receives ${{y}_{{e}}}(n)$ via a linear network as follows:
\vspace{-11mm}
\begin{align}
\mathbf{y}_{r}(n)=\mathbf{H}_{r}\mathbf{s}(n),
\hspace{3mm}{y}_{e}(n)=\mathbf{h}_{e}^{T}\mathbf{s}(n),
\label{noise-formula}
\end{align}
where $\mathbf{H}_{r} \in \mathbb{F}_q^{K\times L}$ represents the network transfer matrix, and $\mathbf{h}_{e} \in \mathbb{F}_q^{L\times 1}$ represents the channel of eavesdropper, which are considered to be unchanged during the delivery phase. This is a valid assumption in most practical scenarios \cite{Shariatpanahi_2016,Multi-antenna-coded-caching,Physical-Layer Schemes}, since changes in the network transfer matrix are due to topology changes such as failure of a node. In this work, we assume that elements of $\mathbf{H}_{r}$, and $\mathbf{h}_{e}$ are i.i.d random variables, and they are chosen uniformly at random from a finite field $\mathbb{F}_q$\cite{Shariatpanahi_2016,Multi-antenna-coded-caching}. Moreover, we suppose large enough $q=2^m$ to ensure that $\mathbf{H}_{r}$, and $\mathbf{h}_{e}$ have the full rank matrix attributes with the high probability.

In the delivery phase (during high-peak hours), each end-user requests an arbitrary file from the library. The requested file of the $i$-th user is denoted by ${{W}_{{{d}_{i}}}}$, and therefore, the demand vector of users is represented by $\mathbf{d}=\{d_1,...,d_K\}$. According to these demands, the transmitters send multi-cast coded messages $\mathbf{{x}_{({{d}_{1}},...,{{d}_{K}})}}$ (which will be defined later) to fulfill the users' request.

We also assume that there exists a totally passive eavesdropper\footnote{This model can be generalized to the case of multiple randomly-located eavesdroppers \cite{multi-eav}, \cite{multi-eav-conference} which will be investigated in our future works.} in the system who observes the multi-cast messages during the delivery phase. It means that the communication happens in an insecure environment. The totally passive eavesdropper can capture, and analyze the secure coded messages, transmitted over a linear network in the second phase. In this work, we assume the eavesdropper does not exist in the placement phase, and thus, we only consider the security of the delivery phase\cite{Secure Delivery}.

For the security requirement, ${{y}_{e}}$ must not reveal any information about the original files $({{W}_{1}},...,{{W}_{N}})$ to the eavesdropper i.e., $I({y}_{e};{{W}_{1}},...,{{W}_{N}})=0$. We show this for our centralized proposed scheme in subsection III-A and for the decentralized scenario in Appendix D, subsection B.

The main performance metric of this paper is to prevent a totally passive eavesdropper from obtaining any information about the original files, $({{W}_{1}},...,{{W}_{N}})$, during the delivery phase. We define the number of required time slots to satisfy a certain demand vector $\mathbf{d}$ by $T_C({\mathbf{d}}) \scriptstyle$. Then, the coding delay will be:
\vspace{-3mm}
\begin{align}
T=\max_{\mathbf{d}} T_C({\mathbf{d}}),
\end{align}
while satisfying the security constraint as:
\vspace{-1mm}
\begin{align}
I({y}_{e};{{W}_{1}},...,{{W}_{N}})=0.
\end{align}
\section{Centralized Multi–Transmitter Coded Caching with Secure Delivery}
\vspace{-1mm}
In this section, we present three different secure centralized multi-transmitter coded caching schemes and analyze their performance. We formulate the secure coding delay, and the closed-form secure key storage expression of the proposed schemes.
\vspace{-2mm}
\subsection{Secure Multi-Transmitter Coded Caching}
\vspace{-2mm}
In this subsection, we propose a secure multi-transmitter caching strategy which is based on the pioneering work \cite{Shariatpanahi_2016} that we integrate the security feature to reach the secure version of it.

\begin{remark}
	In the case of conventional secure scheme, each end-user caches only one unique key per user, and benefits from only the local caching gain provided by encrypted uni-cast delivery. Since there is no shared keys, each transmission is useful to only one user. Thus, the coding delay can be given by $K(1-\frac{M-1}{N-1})$.
\end{remark}

\begin{thm}\label{Theorem_Multi-Server}
	For a linear network with $L$ transmitters, and $K$ cache-enabled users, the following coding delay is securely achievable:
	\begin{align}
	{T}_{S}^{C}=\frac{K(1-\frac{M-1}{N-1})}{L+\frac{K(M-1)}{N-1}},
	\label{eq_coding_delay_multi_transmitter}
	\end{align}
	such that; $M={{M}_{D}}+{{M}_{K}}$, $t=K(\frac{M-1}{N-1})$ and
	\begin{align}
	{{M}_{K}}=\frac{K-t}{K}={{M}_{K}}\{L=1\}, {{M}_{D}}=\frac{Nt}{K},
	\label{eq_M_D_Multi_Transmitter}
	\end{align}
	where ${{M}_{K}}\{L=1\}$ is the key storage in the single-stream scenario \cite{Secure Delivery}, and ${M}_{D}$ shows the data storage. 
	\vspace{-3mm}
	\begin{proof}
		Please refer to appendix A.
	\end{proof}
\end{thm}
\vspace{-4mm}
\begin{remark}
	In the above theorem, ${M}_{K}$ shows the proportion of the users' memory that is occupied by the keys, and provides the system security. Also, the result in \cite{Shariatpanahi_2016} is a special case of our scheme in Algorithm 1 if one does not take the security guarantee into account. Additionally, this formula reduces to the case of \cite{Secure Delivery} when the parameter $L$ is set to one.
\end{remark}

In the rest of this subsection, we present the prefetching and content delivery phases of the proposed scheme. Then, we analyze the performance of the scheme.

\textbf{Prefetching Strategy}: The cache content placement procedure is performed based on \cite{Maddah_2014}. Define $t\triangleq K(\frac{M-1}{N-1})$ and break each file into ${\left( \begin{smallmatrix}
	K \\ 
	t \\ 
	\end{smallmatrix} \right)}$ non-overlapping equal-sized sub-files as:
\begin{align}
{{W}_{n}}=(W_{n,\tau };\tau \subseteq \{1,2,...,K\},\left| \tau  \right|=t), n=1,...,N,
\label{eq_W_Multi_server}
\end{align}
where each sub-file consists of $\frac{f}{\left( \begin{smallmatrix}
	 K \\ 
	 t \\ 
	\end{smallmatrix} \right)}$ symbols.

We further divide each sub-file $W_{n,\tau }$ into $\left( \begin{smallmatrix}
K-t-1 \\ 
L-1 \\ 
\end{smallmatrix} \right)$ non-overlapping equal-sized mini-files:
\vspace{-2mm}
\begin{align}
W_{n,\tau }=\left(W_{n,\tau }^{j}:j=1,..., \left(\begin{smallmatrix}
K-t-1 \\ 
L-1 \\ 
\end{smallmatrix}\right)\right),
\label{eq_mini_file}
\end{align}
where each mini-file consists of $\frac{f}{\left(\begin{smallmatrix}
	K \\ 
	t \\ 
	\end{smallmatrix}\right)\left(\begin{smallmatrix}
	K-t-1 \\ 
	L-1 \\ 
	\end{smallmatrix}\right)}$ symbols.
\vspace{0.1mm}

 Then, the transmitters produce keys ${{K}_{{{\tau }_{k}}}^{(\beta)}}$ using the one-time-pad scheme\cite{one_time_pad} \emph{uniformly} at random as follows:
 \vspace{-1mm}
\begin{align}
Generate \hspace{1mm} {{K}_{{{\tau }_{k}}}^{(\beta)}};&{{\tau }_{k}}\subseteq \left\{ 1,2,...,K \right\},\left| {{\tau }_{k}} \right|=t+L,\beta=1,...,\left( \begin{smallmatrix}
 t+L-1 \\ 
 t \\ 
\end{smallmatrix} \right),
\label{eq_key_generation_multi_server}
\end{align}
where each key consists of $\frac{f}{\left(\begin{smallmatrix}
	K \\ 
	t \\ 
	\end{smallmatrix}\right)\left(\begin{smallmatrix}
	K-t-1 \\ 
	L-1 \\ 
	\end{smallmatrix}\right)}$ symbols.
%\vspace{0.01mm}

Sub-file $W_{n,\tau }$ and key ${{K}_{{{\tau }_{k}}}^{(\beta)}}$ are placed in the cache of user $k$ if $k\in \tau$ and $k\in {\tau}_{k}$, respectively. In this phase, each user must store some random keys to satisfy the security constraint. The transmitters utilize these keys to keep the multi-cast messages safe during the delivery phase.

\textbf{Delivery Strategy}: Consider each ($t+L$)-subset of users which we call $S$. For this specific subset, denote all possible ($t+1$)-subsets of $S$ by ${T}_{i}:i=1,...,\left(\begin{smallmatrix}
 t+L \\ 
 t+1 \\ 
\end{smallmatrix}\right)$. We utilize a $L$-by-$1$ vector $\mathbf{u}_{S}^{{{T}_{i}}}$ for each ${T}_{i}$ to proceed with the zero-forcing procedure as:
\begin{align}
\mathbf{u}_{S}^{{{T}_{i}}}\bot \hspace{1mm} {\mathbf{h}_{j}} \hspace{1.5mm} for \hspace{1mm} all \hspace{1mm} j\in S\backslash {{T}_{i}},\hspace{3mm}
\mathbf{u}_{S}^{{{T}_{i}}} \notperp \hspace{1mm} {\mathbf{h}_{j}} \hspace{1.5mm} for \hspace{1mm} all \hspace{1mm} j\in {{T}_{i}},
\end{align}
while the aforementioned equations can be solved directly \cite{Shariatpanahi_2016,Multi-antenna-coded-caching}.

For each ${T}_{i}$, we define $G({{T}_{i}})$ as an operator to extract all mini-files:
\begin{align}
G({{T}_{i}})={{\varphi }_{r\in {{T}_{i}}}}(W_{{{d}_{r}},{{T}_{i}}\backslash \{r\}}^{j}),
\end{align}
where $W_{{{d}_{r}},{{T}_{i}}\backslash \{r\}}^{j}$ is a mini-file which is accessible within the cache of all clients in ${T}_{i}$, except $r$, and is required by client $r$. In addition, ${\varphi }_{r\in {{T}_{i}}}$ presents a random linear combination of the corresponding mini-files for all $r\in {T}_{i}$. We choose index $j$ carefully such that a specific mini-file does not appear in the former ($t+L$)-subsets. In order to define a new mini-file, index $N(r,T\backslash \{r\})$ can be used as a fresh fragment of file ${{W}_{{{d}_{r}}}}$, which is beneficial to user $r$, and is available in the  cache of users ${T}\backslash \{r\}$, thus we rewrite the previous relation as follows:
\begin{align}
G({{T}_{i}})={{\varphi }_{r\in {{T}_{i}}}}(W_{{{d}_{r,{{T}_{i}}\backslash \{r\}}}}^{N(r,{{T}_{i}}\backslash \{r\})}).
\label{eq-linear-combs}
\end{align}

Ultimately, we define $\mathbf{x}(S)$ as a secure coded message which is assigned to a certain ($t+L$)-subset of users (i.e. $S$):
\vspace{-3mm}
\begin{align}
&\mathbf{x}(S)=\sum\nolimits_{T\subseteq S,\left| T \right|=t+1}({\mathbf{u}_{S}^{T}}G(T))\oplus{K}_{{\tau}_{k}}{\mathbf{w}},
\end{align}
where ${K}_{{\tau}_{k}}$ is the corresponding key, and ${\mathbf{w}}$ is a $L$-by-$1$ vector of ones. Note that in this scheme, we put the same keys on $L$ transmitters at each time slot. Since there is a single-antenna eavesdropper in the delivery phase, he/she is unable to find any useful information about the original files. The benefit of this work is that the key storage will reduce by the factor of $L$. In the rest of this subsection, We will prove the system's security.

Each secure coded message must be repeated with different random coefficients $\left( \begin{smallmatrix}
t+L-1 \\ 
t \\ 
\end{smallmatrix} \right)$ times for a specific ($t+L$)-subset of users to be decodable by all users of $S$. In order to generate different linear combinations of such mini-file, we use the notation of ${\mathbf{x}_{\omega }}(S):\omega =1,...,\left( \begin{smallmatrix}
 t+L-1 \\ 
 t \\ 
\end{smallmatrix} \right)$. Note that the only difference between ${\mathbf{x}_{\omega }}(S)$'s is using random coefficients to make various independent random linear combinations of a certain mini-file with the high probability. Thus, to distinguish between different versions of the above notation, we define:
\begin{align}
{\mathbf{x}_{\omega }}(S)=\sum\nolimits_{T\subseteq S,\left| T \right|=t+1}({\mathbf{u}_{S}^{T}{{G}_{\omega }}(T)})\oplus {{K}_{{\tau}_{k}}^{(\beta)}}{\mathbf{w}},\hspace{3mm}
{{G}_{\omega }}({{T}_{i}})=\varphi _{r\in {{T}_{i}}}^{\omega }(W_{{{d}_{r}},{{T}_{i}}\backslash \{r\}}^{N(r,{{T}_{i}}\backslash \{r\})}).
\end{align}

Eventually, the transmitters send the following block including $\left( \begin{smallmatrix}
t+L-1 \\ 
t \\ 
\end{smallmatrix} \right)$ coded messages corresponding to a certain ($t+L$)-subset of users:
\vspace{-3mm}
\begin{align}
[{\mathbf{x}_{1}}(S),...,{\mathbf{x}_{\left( \begin{smallmatrix} 
		t+L-1 \\ 
		t 
		\end{smallmatrix} \right)}}(S)],
\end{align}
the index $N(r,{{T}}\backslash \{r\})$ will be updated for those mini-files which have appeared in (\ref{eq-linear-combs}). The same procedure will be held to extract all different ($t+L$)-subsets of [$K$].

Algorithm \ref{alg:SECURE-MULTI-SEREVR} demonstrates the pseudo-code of our placement and delivery phases carefully. Let's take a look at the security analysis of our scheme.

\textbf{Security Analysis}:
We remark that the main idea for safeguarding the security in this paper is utilizing the secret shared keys for the multi-transmitter coded caching schemes with $DoF\footnote{It means the number of end-users that can benefit from each coded message at each time slot.}=t+L$. We assume that the transmitters perform the key distribution and management\footnote{Security can be further reinforced by employing the wireless secret key agreement protocols in which the clients, and server agree on secure keys by leveraging the peer-to-peer (P2P) wireless connections\cite{Deep learning}, and \cite{Hardware-Impaired MITM}.}. Each legitimate user can easily decode $\mathbf{x}$ from $\mathbf{y}_{{r}_{i}},i \in [K]$ whereas the eavesdropper fails to unlock the secure coded messages from its received signal because he/she has not cached the keys in the first phase.

 More formally, we define the following \emph{Markov chain} to describe the data processing: (Time index ($n$) is neglected for the sake of notation brevity)
 \vspace{-3mm}
\begin{align}
W \hspace{1mm}\text{---} \hspace{1mm} \mathbf{x} \hspace{1mm} \text{---} \hspace{1mm} {y}_{e}.
\label{markov_chain}
\end{align}
%According to the Markov chain presented in (\ref{markov_chain}), one can infer that we have $I({X};{W}) \ge I({y}_{e};{W})$. In what follows, we consider the pessimistic situation that the eavesdropper obtains the coded messages $X$ from its
%received signal ${y}_{e}$ and prove that we have $I({X};{W})=0$. As a result of the proof, we will also have $I({y}_{e};{W})=0$ as the mutual information metric is non-negative.
\begin{prop}\textit{Secrecy against eavesdropping}:
\vspace{-1mm}
The proposed secure multi-transmitter coded caching scheme achieves the same DoF as known multi-transmitter coded caching strategies, but with secure delivery in the presence of a totally passive eavesdropper, at the cost of more storage. We show that the communication in the delivery phase does not reveal any information to the eavesdropper because he/she has not cached the keys in placement phase. Mathematically, we have:
\vspace{-6mm}
\begin{align}
 I({y}_{e};{{W}_{1}},...,{{W}_{N}})=0.
\end{align}
	\vspace{-1mm}
	\textit{Proof}. To elaborate, we show that the delivery phase does not reveal any information to the eavesdropper. The primary concept behind the proof of security is that the eavesdropper has not cached the keys to be able to unlock the secure coded messages. At the same time, legitimate clients can easily reach the coded messages with the help of their cached keys in the placement phase, without error. We show that:
	\vspace{-5mm}
	\begin{gather}
	\begin{align}
	I\left({y}_{e};{{W}_{1}},...,{{W}_{N}}\right)=0.
	\end{align}
	\end{gather}
	\vspace{-6mm}
	We have,
	\begin{align}
	&I\left({y}_{e};{{W}_{1}},...,{{W}_{N}}\right)=H\left({y}_{e}\right)-H({y}_{e}|{{W}_{1}},...,{{W}_{N}}) \\
	&=H\left({y}_{e}\right)-H(\mathbf{h}_{e}.{\mathbf{x}_{\omega }}\left(S\right):S\subseteq [K],\left| S \right|=t+L,\omega=1,...,\left( \begin{smallmatrix}
	t+L-1 \\ 
	t \\ 
	\end{smallmatrix} \right)|{{W}_{1}},...,{{W}_{N}}) \label{eq_substituting_X}\\
	&=H\left({y}_{e}\right)-H(\mathbf{h}_{e}.\left(\sum\nolimits_{T\subseteq S,\left| T \right|=t+1}({\mathbf{u}_{S}^{T}{{G}_{\omega }}(T)})\oplus {{K}_{S}^{(\beta)}}{\mathbf{w}}\right):S\subseteq [K],\left| S \right|=t+L\nonumber\\ &,\omega=1,...,\left( \begin{smallmatrix}
	t+L-1 \\ 
	t \\ 
	\end{smallmatrix} \right)|{{W}_{1}},...,{{W}_{N}})\label{eq_x_w}\\
	&=H({y}_{e})-H(\mathbf{h}_{e}.(\sum\nolimits_{T\subseteq S,\left| T \right|=t+1}({\mathbf{u}_{S}^{T}}\varphi _{r\in {T}_{i}}^{\omega }(W_{{{d}_{r}},{{T}_{i}}\backslash \{r\}}^{N(r,{{T}_{i}}\backslash \{r\})}))\oplus {{K}_{S}^{(\beta)}}{\mathbf{w}})
	:S\subseteq [K],\left| S \right|=t+L\nonumber\\
	&,\omega=1,...,\left( \begin{smallmatrix}
	t+L-1 \\ 
	t \\ 
	\end{smallmatrix} \right)|{{W}_{1}},...,{{W}_{N}})\label{eq_G_w}\\
	&=H({y}_{e})-H({K}_{S}^{(\beta)} :S\subseteq [K],\left| S \right|=t+L,\omega=1,...,\left( \begin{smallmatrix}
	t+L-1 \\ 
	t \\ 
	\end{smallmatrix} \right)|{{W}_{1}},...,{{W}_{N}})\label{eq_keys}\\
	&=H({y}_{e})-H({K}_{S}^{(\beta)} :S\subseteq [K],\left| S \right|=t+L,\omega=1,...,\left( \begin{smallmatrix}
	t+L-1 \\ 
	t 
	\end{smallmatrix} \right)),\label{eq_substitution}
	\end{align}
	where equation \eqref{eq_x_w} results from substituting the coded message (${\mathbf{x}_{\omega }}\left(S\right)$) in equation \eqref{eq_substituting_X}. One can reach equation \eqref{eq_G_w} by replacing ${{{G}_{\omega }}(T)}$ in equation \eqref{eq_x_w}.
	By assuming that the statistics of the eavesdropper channel ($\mathbf{h}_{e}$) is known \footnote{The pessimistic case of unknown channels will be studied in the future \cite{multi-eav,multi-eav-conference}.} \cite{multi-eav,multi-eav-conference,Lightweight-channel}, and the $F$-bit discrete files are represented by $m$-bit symbols over finite field GF($q$), and considering the fact that beam-forming vectors of users (${\mathbf{u}_{S}^{T}}$) are publicly-known variables, one can reach equation \eqref{eq_keys} from equation \eqref{eq_G_w} . This way, the only term that provides randomness in the form of a key is, ${{K}_{S}^{(\beta)}}$ which occupies a part of the memory of each user as security cost. In other words, keys ${{K}_{S}^{(\beta)}}$ guarantee the system's security. Furthermore, the last equality follows from the fact that the keys are uniformly distributed and are independent of the files $\left({{W}_{1}},...,{{W}_{N}}\right)$. Using the fact that $H\left(A,B\right)\le H\left(A\right)+H\left(B\right)$. We have:
	\vspace{-3mm}
	\begin{align}
	& H({y}_{e})=H(\mathbf{h}_{e}.\Bigl(\sum\nolimits_{T\subseteq S,\left| T \right|=t+1}({\mathbf{u}_{S}^{T}}\varphi _{r\in {{T}_{i}}}^{\omega }(W_{{{d}_{r}},{{T}_{i}}\backslash \{r\}}^{N(r,{{T}_{i}}\backslash \{r\})}))\nonumber\\
	&\oplus {{K}_{{S}}^{(\beta)}}{\mathbf{w}}\Bigl):{S}\subseteq [K],\left| {S} \right|=t+L,\omega=1,...,\left( \begin{smallmatrix}
	t+L-1 \\ 
	t \\ 
	\end{smallmatrix} \right))\\
	& \le \sum\nolimits_{i=1}^{\left( \begin{smallmatrix} 
		K \\ 
		t+L 
		\end{smallmatrix} \right)\left( \begin{smallmatrix} 
		t+L-1 \\ 
		t 
		\end{smallmatrix} \right)}H(\mathbf{h}_{e}.\Bigl(\sum\nolimits_{T\subseteq S_i,\left| T \right|=t+1}({\mathbf{u}_{{S}_{i}}^{T}}\varphi _{r\in {{T}_{i}}}^{\omega }(W_{{{d}_{r}},{{T}_{i}}\backslash \{r\}}^{N(r,{{T}_{i}}\backslash \{r\})}))\oplus {{K}_{{S}_{i}}^{(\beta)}}{\mathbf{w}}\Bigl)\nonumber\\
	&:{S}_{i}\subseteq [K],\left| {S}_{i} \right|=t+L,\omega=1,...,\left( \begin{smallmatrix}
	t+L-1 \\ 
	t \\ 
	\end{smallmatrix} \right))= \frac{\left(\begin{smallmatrix}
		K \\ 
		t+L \\ 
		\end{smallmatrix}\right) \left(\begin{smallmatrix}
		t+L-1 \\ 
		t \\ 
		\end{smallmatrix}\right)}{{{\log }_{2}}\left(\frac{F}{\left(\begin{smallmatrix}
			K \\ 
			t \\ 
			\end{smallmatrix}\right) \left(\begin{smallmatrix}
			K-t-1 \\ 
			L-1 \\ 
			\end{smallmatrix}\right)}\right)}.\label{eq_entropy_keys}
	\end{align}

	On the other hand, we have:
	\vspace{-5mm}
	\begin{align}
	& H({K}_{{S}}^{(\beta)} :{S}\subseteq [K],\left| {S} \right|=t+L,\omega=1,...,\left( \begin{smallmatrix}
	t+L-1 \\ 
	t \\ 
	\end{smallmatrix} \right))\nonumber\\
	&=\sum\nolimits_{i=1}^{ \left(\begin{smallmatrix} 
		K \\ 
		t+L 
		\end{smallmatrix}\right)  \left(\begin{smallmatrix} 
		t+L-1 \\ 
		t 
		\end{smallmatrix}\right)}H({K}_{{S}_{i}}^{(\beta)} :{S}_{i}\subseteq [K],\left| {S}_{i} \right|=t+L,\omega=1,...,\left( \begin{smallmatrix}
	t+L-1 \\ 
	t 
	\end{smallmatrix} \right))=\frac{ \left(\begin{smallmatrix}
		K \\ 
		t+L 
		\end{smallmatrix}\right)  \left(\begin{smallmatrix}
		t+L-1 \\ 
		t  
		\end{smallmatrix}\right)}{{{\log }_{2}}\left(\frac{F}{ \left(\begin{smallmatrix}
			K \\ 
			t  
			\end{smallmatrix}\right)   \left(\begin{smallmatrix}
			K-t-1 \\ 
			L-1 
			\end{smallmatrix}\right)}\right)},\label{eq_independent_keys}
	\end{align}
	where the equality in \eqref{eq_independent_keys} demonstrates that all of the keys ${K}_{{S}_{i}}^{(\beta)}$ are generated uniformly at random and mutually independent. By substituting \eqref{eq_entropy_keys} and \eqref{eq_independent_keys} into \eqref{eq_substitution} and considering the fact that for any $X$ and $Y$, $I\left({X};{Y}\right) \ge 0$, we have:
	\begin{align}
	I\left({y}_{e};{{W}_{1}},...,{{W}_{N}}\right)=0.
	\end{align}
\end{prop}
\vspace{-2mm}
Numerical evaluations will show that the cost of security in the system is negligible in terms of memory usage when the number of files and users increase.

\begin{table*}[t]
	\centering
	\caption{Comparison of Different Schemes. ($d$ stands for $\min(s+L-1,K)$, and TX column represents the number of transmitters.) }
	\begin{tabular}{ |c|c|c|c|c|c|c|c| } 
		\hline
		{\scriptsize{\textbf{Scheme}}} & Coding Delay & ${M}_{K}$ & {\scriptsize{DoF}} & TX & {\scriptsize{Security}}  & {\scriptsize{Subpacketization}}\\ 
		\hline
		{\scriptsize{Centralized \cite{Secure Delivery}}} & $\dfrac{K(1-\dfrac{M-1}{N-1})}{1+\dfrac{K(M-1)}{N-1}}$ & $1-\frac{t}{K}$ & $t+1$  & $1$ & $\checkmark$  & $\frac{1}{\left(\begin{smallmatrix}
			K \\ 
			t \\ 
			\end{smallmatrix}\right)}$ \\ 
		\hline
		{\scriptsize{Scheme-Th. 1}} & $\dfrac{K(1-\dfrac{M-1}{N-1})}{L+\dfrac{K(M-1)}{N-1}}$ & $1-\frac{t}{K}$ & $t+L$  & $L$ & $\checkmark$  & $\frac{1}{\left(\begin{smallmatrix}
			K \\ 
			t \\ 
			\end{smallmatrix}\right)\left(\begin{smallmatrix}
			K-t-1 \\ 
			L-1 \\ 
			\end{smallmatrix}\right)}$ \\
		\hline
		\cite{Shariatpanahi_2016} & $\dfrac{K(1-\dfrac{M}{N})}{L+\dfrac{KM}{N}}$ & --- & $t+L$  & $L$ & \large{$\times$}  & $\frac{1}{\left(\begin{smallmatrix}
			K \\ 
			t \\ 
			\end{smallmatrix}\right)\left(\begin{smallmatrix}
			K-t-1 \\ 
			L-1 \\ 
			\end{smallmatrix}\right)}$ \\
		\hline
		{\scriptsize{Scheme-Th. 2}} & $\dfrac{K(1-\dfrac{M-L}{N-L})}{L+\dfrac{K(M-L)}{N-L}}$ & $L(1-\frac{t}{K})$ & $t+L$  & $L$ & $\checkmark$ & $\frac{1}{\left(\begin{smallmatrix}
			\sfrac{K}{L} \\ 
			\sfrac{t}{L} \\ 
			\end{smallmatrix}\right)}$  \\
		\hline
		\cite{Adding_transmitters} & $\dfrac{K(1-\dfrac{M}{N})}{L+\dfrac{KM}{N}}$ & --- & $t+L$  & $L$ & \large{$\times$} & $\frac{1}{\left(\begin{smallmatrix}
			\sfrac{K}{L} \\ 
			\sfrac{t}{L} \\ 
			\end{smallmatrix}\right)}$  \\
		\hline
		{\scriptsize{Scheme-Th. 3}} & $\dfrac{K-t}{L+t}$ & $\frac{L(K-t)(t+1)}{K(t+L)}$  & $t+L$ & $L$ & $\checkmark$  &$\frac{1}{\left(\begin{smallmatrix}
			K \\ 
			t \\ 
			\end{smallmatrix}\right) \left(\begin{smallmatrix}
			K-t-1 \\ 
			L-1 \\ 
			\end{smallmatrix}\right)(L+t)}$  \\
		\hline
		\cite{Feedback-bottleneck} & $\dfrac{K-t}{L+t}$ & ---  & $t+L$ & $L$ &  \large{$\times$}  &$\frac{1}{\left(\begin{smallmatrix}
			K \\ 
			t \\ 
			\end{smallmatrix}\right) \left(\begin{smallmatrix}
			K-t-1 \\ 
			L-1 \\ 
			\end{smallmatrix}\right)(L+t)}$  \\
		\hline
		{\scriptsize{Decentralized \cite{Secure Delivery}}} & \parbox{3mm}{\begin{align*}
			&\min \{\frac{N-1}{K(M-1)}.(1-{{(1-\frac{M-1}{N-1})}^{K}})\\
			&,1\}\times K(1-\frac{M-1}{N-1})\end{align*}}& $1-q$ & $s$  & 1 & $\checkmark$  & $\frac{1}{\left(\begin{smallmatrix}
			K \\ 
			s \\ 
			\end{smallmatrix}\right)}$ \\ 
		\hline
		{\scriptsize{Scheme-Th. 4}} & $\sum\limits_{s=1}^{K}{\frac{\left( \begin{smallmatrix}
				K \\ 
				d \\ 
				\end{smallmatrix} \right){{(q)}^{s-1}}{{(1-q)}^{K-s+1}}}{\left( \begin{smallmatrix}
				K-s \\ 
				d-s \\ 
				\end{smallmatrix} \right)}}
		\times	\left( \begin{smallmatrix}
		d-1 \\ 
		s-1 \\ \nonumber
		\end{smallmatrix} \right)$ & \parbox{3mm}{ \begin{align*}
			&\sum\limits_{s=1}^{K}{\frac{\left( \begin{smallmatrix}
					K-1 \\ 
					d-1 \\ 
					\end{smallmatrix} \right)\left( \begin{smallmatrix}
					d-1 \\ 
					s-1 \\ 
					\end{smallmatrix} \right)}{\left( \begin{smallmatrix}
					K-s \\ 
					d-s \\ 
					\end{smallmatrix} \right)}}\times\\
				& {{\left( q \right)}^{s-1}}{{\left( 1-q \right)}^{K-s+1}}\end{align*}} & $d$  & $L$ & $\checkmark$ & $\frac{1}{\left(\begin{smallmatrix}
			K \\ 
			s \\ 
			\end{smallmatrix}\right)\left(\begin{smallmatrix}
			K-s \\ 
			L-1 \\ 
			\end{smallmatrix}\right)}$  \\
		\hline
		\cite{WSA2021,Decentralized_cyclic} & $\sum\limits_{s=1}^{K}{\frac{\left( \begin{smallmatrix}
				K \\ 
				d \\ 
				\end{smallmatrix} \right){{(q)}^{s-1}}{{(1-q)}^{K-s+1}}}{\left( \begin{smallmatrix}
				K-s \\ 
				d-s \\ 
				\end{smallmatrix} \right)}}
		\times	\left( \begin{smallmatrix}
		d-1 \\ 
		s-1 \\ \nonumber
		\end{smallmatrix} \right)$ & --- & $d$  & $L$ & \large{$\times$} & $\frac{1}{\left(\begin{smallmatrix}
			K \\ 
			s \\ 
			\end{smallmatrix}\right)\left(\begin{smallmatrix}
			K-s \\ 
			L-1 \\ 
			\end{smallmatrix}\right)}$  \\
		\hline
	\end{tabular}
	\vspace{-0.2mm}
	%{\raggedright Note: In this table, for convenience, $d$ stands for $\min(s+L-1,K)$. Also, ${N}_{T}$ represents the number of transmitters.}
	\label{TABLE_2}
\end{table*}

\begin{remark}
Theorem \ref{Theorem_Multi-Server} results from assigning each key to each subset of $t+L$ users. By assigning each key to each subset of $t+1$ users, the key storage will be modified as:
\begin{align}
{{M}_{K}}=\frac{K-t}{K}\left(\begin{smallmatrix}
 t+L-1 \\ 
 t \\ 
\end{smallmatrix}\right),
\end{align} 
which is clearly worse in terms of memory usage.
\end{remark}

\begin{algorithm}[ht!]
	
	\caption{Secure Multi-Transmitter Coded Caching Algorithm }\label{alg:SECURE-MULTI-SEREVR}
	
	\begin{algorithmic}[1]
		\scriptsize
		\STATE \textbf{Centralized Cache Placement $({{W}_{1}},...,{{W}_{N}} )$ }
		\STATE $t=K(\frac{M-1}{N-1})$
		\FOR{\textbf{all}  $ n \in \{ 1,2,...,N\} $}
		\STATE Split file ${{W}_{n}}$ into equal-sized sub-files $W_{n,\tau };\tau \subseteq \{1,2,...,K\},\left| \tau  \right|=t$
		\FOR {\textbf{all} $W_{n,\tau };\tau \subseteq \{1,2,...,K\},\left| \tau  \right|=t$ }
		\STATE Split $W_{n,\tau }$ into $\left(W_{n,\tau }^{(j)}:j=1,..., \left(\begin{smallmatrix}
		 K-t-1 \\ 
		 L-1 \\ 
		\end{smallmatrix}\right)\right)
		$ of equal-sized mini-files.
		\ENDFOR
		\ENDFOR
		\STATE \textbf{Generate Keys ${{K}_{{{\tau }_{k}}}^{(\beta)}}$ such that ${{\tau }_{k}} \subseteq \{1,2,...,K\},\left| {{\tau }_{k}}  \right|=t+L, \beta=1,...,\left( \begin{smallmatrix}
			 t+L-1 \\ 
			 t \\ 
			\end{smallmatrix} \right)$}
		\FOR{$ k \in \{ 1,2,...,K\} $}
		\FOR{$n$=1,2,...,$N$}
		\STATE File $W_{n,\tau }^{(j)}$ is placed in cache, ${{Z }_{k}}$, of user $k$ if $k \in {\tau }$
		\STATE Key $K_{{{\tau }_{k}}}^{(\beta)}$ is placed in cache, ${{Z }_{k}}$, of user $k$ if $k \in {\tau }_{k}$
		\ENDFOR
		
		\ENDFOR		
		\STATE \textbf{Content Delivery $(d_1 ,... , d_K )$ }
		\FOR {\textbf{all} $T\subseteq [K],\left| T \right|=t+1$}
		\FOR {\textbf{all} $r\in T$}
		\STATE $N(r,T\backslash \{r\})\leftarrow 1$  
		\ENDFOR
		\ENDFOR
		\FOR {\textbf{all} $S\subseteq [K],\left| S \right|=t+L$}
		\FOR {\textbf{all} $T\subseteq S,\left| T \right|=t+1$}
		\STATE Design $\mathbf{u}_{S}^{T}$ such that: for all $j\in S$, ${\mathbf{h}_{j}}\perp \mathbf{u}_{S}^{T}$ if $j \notin T$ and ${\mathbf{h}_{j}}\notperp \mathbf{u}_{S}^{T}$ if $j \in T$
		\ENDFOR
		\FOR {\textbf{all} $\omega =1,...,\left(\begin{smallmatrix}
			 t+L-1 \\ 
			 t \\ 
			\end{smallmatrix}\right)$}
		\FOR {\textbf{all} $T\subseteq S$, $\left| T \right|=t+1$ }
		\STATE ${{G}_{\omega }}(T)\leftarrow \varphi _{r\in T}^{\omega }(W_{{{d}_{r}},T\backslash \{r\}}^{N(r,T\backslash \{r\})})$ 
		\ENDFOR
		\STATE ${\mathbf{x}_{\omega }}(S)\leftarrow \sum\limits_{T\subseteq S,\left| T \right|=t+1}({\mathbf{u}_{S}^{T}{{G}_{\omega }}}(T))\oplus {{K}_{{\tau}_{k}}^{(\beta)}}{\mathbf{w}}$
		\ENDFOR
		\STATE transmit $\mathbf{X}(S)=[{\mathbf{x}_{1}}(S),...,{\mathbf{x}_{\left(\begin{smallmatrix} 
				t+L-1 \\ 
				t 
				\end{smallmatrix}\right)}}(S)]$
		\FOR {\textbf{all} $T\subseteq S$, $\left| T \right|=t+1$ }
		\FOR {\textbf{all} $r\in T$}
		\STATE ${N(r,T\backslash \{r\})\leftarrow N(r,T\backslash \{r\})+1}$
		\ENDFOR
		\ENDFOR
		\ENDFOR
		%\FOR {\textbf{all} $r\in T$}
	%	\STATE ${N(r,T\backslash \{r\})\leftarrow N(r,T\backslash \{r\})+1}$
	\end{algorithmic}
\end{algorithm}
\setlength{\textfloatsep}{4pt}
In Table I, we compare different schemes from various viewpoints such as, key storage, coding delay, security, DoF, and subpacketization level.

In the next subsections, we examine two other centralized multi-transmitter caching schemes from the perspective of security integration. Although they perform worse than Algorithm 1 in terms of memory usage i.e., (the security cost in our problem), they have the advantage of reducing subpacketization and feedback, as two major challenges of cache-aided networks.
\vspace{-3mm}
\subsection{Secure Multi-Transmitter Coded Caching with Reduced Subpacketization}
\vspace{-1.5mm}
In the context of cache-aided communication in MISO-BC, ref \cite{Adding_transmitters} reveals that having a multi-transmitter scheme can amazingly reform the long-standing subpacketization challenge of coded caching by reducing the required subpacketization to nearly its $L$-th root. Although the scheme works for all values of $K$, $L$, and $M$, we consider the case when $L|t$ and $L|K$ jointly. In this subsection, we integrate security into \cite{Adding_transmitters} to achieve the secure version of it.

\begin{thm}
	For a cache-aided network with $L$ transmitters, and $K$ cache-enabled receiving users, the following coding delay is securely achievable:
	\begin{align}
	{T}_{S}^{C}=\frac{K(1-\frac{M-L}{N-L})}{L+\frac{K(M-L)}{N-L}},
	\label{eq_delay_adding_transmitters}
	\end{align}
	such that; $M={{M}_{D}}+{{M}_{K}}$, $t=K(\frac{M-L}{N-L})$ and
	\begin{align}
	{{M}_{K}}=L\times (\frac{K-t}{K}), {{M}_{D}}=\frac{Nt}{K},
	\label{eq_M_adding_transmitters}
	\end{align}
	where ${M}_{K}$ represents the key storage size, and guarantees the security of the scheme.
	\begin{proof}
		Please refer to appendix B.
	\end{proof}
\end{thm}
\vspace{-2mm}
In the above Theorem, ${M}_{K}$ indicates the key storage, and provides the system security. This formula will reduce to the case in \cite{Secure Delivery} provided that the parameter $L$ sets to one. Also, \cite{Adding_transmitters} is a particular case of our scheme if one does not take the security guarantee into consideration.

In the rest of this subsection, we present the achievable scheme for the case in which $L|t$ and $L|K$ simultaneously.

\textbf{Prefetching Strategy}: Firstly, we break the $K$ users into ${K^\prime}\triangleq \frac{K}{L}$ disjoint groups as follows:
\vspace{-3mm}
\begin{align}
{G}_{g}=\{l{K^\prime}+g,\hspace{1mm} l=0,1,...,L-1\}, for \hspace{1.5mm} g=1,2,...,{K^\prime},
\label{eq_grouping}
\vspace{-3mm}
\end{align}
where each group ${G}_{g}$ includes $L$ users. We aim to apply algorithm of \cite{Maddah_2014} to serve $\frac{t}{L}+1$ groups. Since each group ${G}_{g}$ contains $L$ members, we serve $t+L$ users simultaneously at each time slot. To this end, we split each file ${W}_{n}$ into $ \left(\begin{smallmatrix}
 \sfrac{K}{L} \\ 
\sfrac{t}{L} \\ 
\end{smallmatrix}\right) $ non-overlapping equal-sized sub-files as:
\begin{align}
	{W}_{n}=\{{{W_{n,\tau}}};\tau \subseteq \{1,2,...,K'\}:\left| \tau  \right|=\frac{t}{L}\}, n=1,...,N.
	\label{eq_fraction}
\end{align}
%Then, we consider each ($\frac{t}{L}+1$)-subset of [$K'$], which we call $S$.

Next, $L$ transmitters generate keys ${{K}_{{{\tau }_{k}}}^{(\beta)}}$, using the one-time-pad scheme \cite{one_time_pad}, uniformly at random as follows:
\vspace{-2mm}
\begin{align}
Generate\hspace{1mm} {{K}_{{{\tau }_{k}}}^{(\beta)}};{\tau }_{k} \subseteq  \{1,2,...,K'\}:\left| {\tau}_{k}  \right|=\frac{t}{L}+1\},\hspace{3mm} \beta =1,...,L.
\label{key_adding_trans}
\end{align}

The subfile ${{W_{n,\tau}}}$ is placed in the cache of user $k$ if $k \in \tau$. Also, key ${{K}_{{{\tau }_{k}}}^{(\beta)}}$ is placed in the cache of all users belonging to groups ${\tau}_{k}$. For example $K_{12}^{(\beta)}$, for $\beta=1,...,L$, is placed in cache of all users of groups $G_1$, and $G_2$. As a result of grouping users, the users in the same group have an identical cache.

\textbf{Delivery Strategy}: We consider each ($\frac{t}{L}+1$)-subset of [$K'$], which we call $S$. After revealing the requests where each user $k$ requires file ${W}_{{d}_{k}}$, ${d}_{k} \in [N]$, the delivery phase consists of a sequential transmission $\mathbf{x}(S)$, where each transmission takes the following form:
\vspace{-2mm}
	\begin{align}
	\mathbf{x}(S)=\sum\nolimits_{g\in S}{\sum\nolimits_{k\in {{G}_{g}}}({W_{{d}_{k},{S\backslash g}}}}\hspace{1mm}{\mathbf{h}^{{{G}_{g}}\backslash k}})\oplus \mathbf{K}_{{{\tau }_{k}}}^{(\beta )},
\end{align}
where $\mathbf{K}_{{{\tau }_{k}}}^{(\beta )}= \left(\begin{smallmatrix}
K_{{{\tau }_{k}}}^{(1)} \\ 
\vdots  \\ 
K_{{{\tau }_{k}}}^{(L)} \\ 
\end{smallmatrix}\right) $ is the corresponding key, and ${\mathbf{h}^{{{G}_{g}}\backslash k}}$
is a $L$-by-$1$ precoding vector that is designed to belonging in the null space of the channel between $L$ transmitters, and $L-1$ users in group ${G}_{g}$, except user $k \in {G}_{g}$.

Decoding Process: Each user $k \in {G}_{g}$ can utilize its cache to remove all undesired sub-files by its own group ${G}_{g}$, it implies that user $k \in {G}_{g}$ can remove
\begin{align}
	\sum\nolimits_{g'\in S\backslash g}{\sum\nolimits_{j\in {{G}_{g'}}}{W_{{d}_{j},{S\backslash g'}}}}\hspace{0.5mm}{\mathbf{h}^{{{G}_{g'}}\backslash j}},
\end{align}
because $g'\ne g\in S$, i.e., because the cache of receiver $k$ includes all sub-files ${W_{{d}_{j},{S\backslash g'}}}$ in the above summation. This allows user $k$ to eliminate all undesired sub-files to its group ${G}_{g}$. To complete the decoding procedure, the interference term for user $k$ can only come from that sub-files that have been requested by $L-1$ other users of its own group ${G}_{g}$. We use zero-forcing procedure to eliminate this interference, and user $k$ can receive the desired sub-file ${W_{{d}_{k},{S\backslash g}}}$. This process will be performed instantly for all user $k \in {G}_{g}$, and for all $g \in S$. Hence, this way, $\frac{t}{L}+1$ groups, and $t+L$ users can benefit from the coded messages at each time slot. This process will be done for all $S \in [K']$. As a result, in this scheme, we achieve the DoF of $t+L$ with the subpacketization of $ \left(\begin{smallmatrix}
\sfrac{K}{L} \\ 
\sfrac{t}{L} \\ 
\end{smallmatrix}\right)$.
\vspace{-3mm}
\subsection{Secure Multi-Transmitter Coded Caching with Reduced Feedback}
\vspace{-3mm}
In this subsection, we present the scheme that is based on the proposed method in \cite{Feedback-bottleneck}, in which we integrated the security guarantee to reach the secure version of it.

\begin{thm}\label{Theorem_Performance}
	In the $K$-user cache-aided network with $L$
	transmitters and cache size $M$, the scheme achieves the following coding delay while satisfying the security constraint:
	\begin{align}
	{T_{S}^{C}}=\frac{K-t}{L+t},
	\label{eq_delay_feedback}
	\end{align}
	such that; $M={{M}_{D}}+{{M}_{K}}$, $t={K{{M}_{D}}}/{N}$ and
	\begin{align}
	{{M}_{K}}=\frac{L(K-t)(t+1)}{K(t+L)}.
	\label{eq_M_feedback}
	\end{align}
	where ${M}_{K}$ and ${M}_{D}$ are the key and data storage, respectively. 
	\begin{proof}
		Please refer to Appendix C.
	\end{proof}
\end{thm}

We draw attention to the fact that in the aforementioned statement, ${M}_{K}$ displays the percentage of the user's memory that is taken up by the keys and offers system security. This formula reduces to the case in \cite{Secure Delivery} when the parameter $L$ is set to one. Furthermore, \cite{Feedback-bottleneck} is a specific case of our scheme if the security feature is ignored.

In the rest of this subsection, we present the proposed scheme’s prefetching and content delivery phases. The system is described for the case where ${{t}}/{L}\in \mathbb{N}$ ($\large{t\triangleq \sfrac{K{{M}_{D}}}{N}}$, where ${M}_{D}$ is data storage), while the remaining regimes are attained using memory sharing with a slight decrease in DoF. 

\textbf{Prefetching Strategy}: The prefetching phase is performed without prior knowledge of actual requests. This strategy remains the same as \cite{Maddah_2014}. Each file ${{W}_{n}}$ is broken into $K \choose t$ non-overlapping equal-sized fragments as follows:
\vspace{-2mm}
\begin{align}
{{W}_{n}}=(W_{n,\tau };\tau \subseteq \{1,2,...,K\},\left| \tau  \right|=t), n=1,...,N ,\label{eq_fragments_files}
\end{align}
Then, the transmitters generate keys ${{K}_{{{\tau }_{k}}}^{(\beta)}}$ by using the one-time-pad scheme\cite{one_time_pad}, \emph{uniformly} at random as follows:\vspace{-2mm}
\begin{align}
Generate \hspace{1mm} {{K}_{{{\tau }_{k}}}^{(\beta)}};{{\tau }_{k}}\subseteq \left\{ 1,2,...,K \right\},\hspace{3mm}\left| {{\tau }_{k}} \right|=t+1,\hspace{3mm}\beta=1,...,\left( \begin{smallmatrix}
K-t-1 \\ 
L-1 \\ 
\end{smallmatrix} \right)(t+1)L. \label{eq_keys_generation_feedback}
\end{align}

Subfile $W_{n,\tau }$, and key ${{K}_{{{\tau }_{k}}}^{(\beta)}}$ are placed in the cache of user $k$ if $k\in \tau$ and $k\in {{\tau }_{k}}$, respectively. During the placement phase, each user must cache some certain random shared keys to fulfill the security constraint. The transmitters use these secret shared keys to keep the multi-cast messages secure during the delivery phase.

\textbf{Delivery Strategy}: In the delivery phase, each of the $K$ users requests one of the $N$ files from the library. The transmitters reply to these requests by sending coded messages that serve $t+L$ users simultaneously in each time slot. Furthermore, the users are divided into two sets, $\lambda \subset \left[ K \right]$, $\left| \lambda \right|=L$, and $\pi \subset \left[ K \right]\backslash \lambda$ , $\left| \pi 
\right|=t$. Then, each sub-file $W_{{d}_{k},\tau }$ is segmented twice:
\vspace{-2mm}
\begin{align}
& W_{{d}_{k},{\tau }}=(W_{{d}_{k},\sigma ,\tau },\sigma \subseteq \left[ K \right]\backslash (\tau \cup \left\{ k \right\}),\left| \sigma  \right|=L-1),\label{eq_key_split1}\\
& W_{{d}_{k},\sigma ,\tau }=(W_{{d}_{k},\sigma ,\tau }^{r},r\in \left[ L+t \right]),\label{eq_key_split2}
\end{align}
\vspace{-2mm}
where $\sigma$, and $r$ are the counter parameters related to the subpacketization procedure.

Next, we describe the details of the delivery scheme presented in Algorithm 2. First, the set $\lambda$ with the size of $L$ users is selected. Then, the zero-forcing procedure is designed at the receiver side. For some sets $\lambda \subset \left[ K \right]$ of $\left| \lambda \right|=L$ end-users, we design $\mathbf{H}_{\lambda }^{-1}$, as the normalized inverse of the $L\times L$ channel matrix ${\mathbf{H}_{\lambda }}$ equivalent to the channel matrix between the transmitters, and end-users of set $\lambda$. More formally, let us define:
\vspace{-2mm}
\begin{align}
{\mathbf{h}_{k}} \hspace{1mm} \bot \hspace{1mm} {\mathbf{h}_{\lambda \backslash \lambda (l)}} \hspace{1.5mm} if \hspace{1.5mm} k\in \lambda \backslash \lambda (l),
\end{align}
where ${\mathbf{h}_{\lambda \backslash \lambda (l)}}$ is the $l$-th column of $\mathbf{H}_{\lambda }^{-1}$, $l\in \left[ L \right]$. Next, another set $\pi \subseteq \left[ K \right]\backslash \lambda$ with the size of $t$ users is chosen, where $\lambda$ and $\pi$ are disjoint. The members of set $\pi$ are partitioned to $L$ distinct subsets by the name of ${{\phi }_{i}}$, each having $\frac{t}{L}$ users.

In the next step, we construct the vector of $L$ coded messages, and use $s$ and ${{v}_{i}}$ as operators to extract all $L$ different sets $\mu$,  for achieving the full multi-cast. In each iteration, Algorithm \ref{alg:SMACC} links a user from $\lambda$ with some set ${{\phi }_{{{v}_{i}}}}$ for the sake of creating set $\mu$, and at the end of the procedure, after $L$ iterations, each user from $\lambda$ will link with each set ${{\phi }_{{{v}_{i}}}}$. Subsequently, we can write the coded message as follows:
\vspace{-2mm}
\begin{align}
X_{\mu }^{v,\sigma }=\underset{k\in \mu }{\mathop{\oplus }}\,W_{{{d}_{k}},\sigma ,(v\cup \mu )\backslash \{k\}}\oplus {{K}_{v\cup \mu }} ,
\end{align}

Each transmitted coded message is useful for $\frac{t}{L}+1$ users, which is specified by set $\mu$. The set ${v}$ is referred to those users who have cached a specific sub-file that is beneficial for users in set $\mu$. It is noticeable that $\mu \cap v=0$. Furthermore, we suppose a set $\sigma \subseteq (\left[ K \right]\backslash (\mu \cup v)),\left| \sigma  \right|=L-1$ which determines the users that the corresponding sub-file is zero-forced to them. Eventually, we have an interference term for the eavesdropper who sniffs the network. The significant difference between our work, and \cite{Feedback-bottleneck} is adding a secret shared key to each coded message leading to more storage as the security cost. However, this additional cost becomes negligible, as shown in the numerical evaluations section. Algorithm 2 demonstrates the pseudo-code of our placement and delivery phases carefully.
\vspace{-4.5mm}
\vspace{-0.1mm}
\begin{algorithm}[ht]

	\caption{Secure Multi-Transmitter Coded Caching with Reduced Feedback}\label{alg:SMACC}

	\begin{algorithmic}[1]
		\small
		\STATE \textbf{Centralized Cache Placement $({{W}_{1}},...,{{W}_{N}} )$ }
		\STATE $t={K{{M}_{D}}}/{N}$
		\FOR{  $ n \in \{ 1,2,...,N\} $}
		\STATE Split file ${{W}_{n}}$ into equal-sized fragments ${W_{n,\tau}};\tau \subseteq \{1,2,...,K\},\left| \tau  \right|=t$
		\ENDFOR
		\STATE \textbf{Generate Keys ${{K}_{{{\tau }_{k}}}^{(\beta)}}$ such that ${{\tau }_{k}} \subseteq \{1,2,...,K\},\left| {{\tau }_{k}}  \right|=t+1, \beta=1,...,\left( \begin{smallmatrix}
			K-t-1 \\ 
			L-1 \\ 
			\end{smallmatrix} \right)(t+1)L$}
		\FOR{$ k \in \{ 1,2,...,K\} $}
		\FOR{$n$=1,2,...,$N$}
		\STATE File ${W_{n,\tau}}$ is placed in cache, ${{Z }_{k}}$, of user $k$ if $k \in {\tau }$
		\STATE Key ${{K}_{{{\tau }_{k}}}^{(\beta)}}$ is placed in cache, ${{Z }_{k}}$, of user $k$ if $k \in {\tau }_{k}$
		
		\ENDFOR
		
		\ENDFOR		
		\STATE \textbf{Content Delivery $(d_1,..., d_K )$ }
		\FOR{$\lambda  \subset \left[ K \right],\left| \lambda  \right|=L$}
		\STATE Calculate $\mathbf{H}_{\lambda }^{-1}$
		\FOR{$\pi \subseteq (\left[ K \right]\backslash \lambda ),\left| \pi  \right|=t$}
		\STATE Break $\pi$ into some ${{\phi }_{i}} \hspace{2mm} i \in \left[ L \right]: \left| {{\phi }_{i}} \right|=\frac{t}{L}$,
		\STATE ${{\bigcup }_{i\in \left[ L \right]}}\hspace{1.25mm} {{\phi }_{i}}=\pi ,{{\phi }_{i}}\cap {{\phi }_{j}}=0,\forall i,j\in \left[ L \right]$
		\FOR{ $ s \in \{ 0,1,...,L-1\} $}
		\STATE ${{v}_{i}}=((s+i-1)\bmod L)+1,i\in \left[ L \right]$
		\STATE Transmit
		\STATE 
		\[{{\mathbf{x}}_{\lambda ,\pi }^{s}}=\mathbf{H}_{\lambda }^{-1}.\left[ \begin{smallmatrix}
		X_{\lambda (1)\cup {{\phi }_{{{v}_{1}}}}}^{\pi \backslash {{\phi }_{{{v}_{1}}}},\lambda \backslash \lambda (1)}\ \oplus {{K}_{\pi \cup \lambda (1)}}  \\
		\vdots   \\
		X_{\lambda (L)\cup {{\phi }_{{{v}_{L}}}}}^{\pi \backslash {{\phi }_{{{v}_{L}}}},\lambda \backslash \lambda (L)} \oplus {{K}_{\pi \cup \lambda (L)}} \\
		\end{smallmatrix} \right]\]
		\ENDFOR
		\ENDFOR
		\ENDFOR	
	\end{algorithmic}
\end{algorithm}
\setlength{\textfloatsep}{8pt}
%In the next section, we show that the multicasting/multiplexing gain can still be attained in a more decentralized setting. Hence, we propose a secure decentralized scheme, in which content placement is performed in a decentralized manner. Although there is a lack of coordination, the proposed scheme is able to generate secure multi-cast coded messages, and achieves the coding delay that is \emph{asymptotically} equal compared to the centralized schemes when the network is scaled up.
\section{Decentralized Multi-Transmitter Coded Caching with Secure Delivery}
\vspace{-3mm}
In many scenarios, central transmitters may not be available and therefore, we propose a secure caching scheme, in which the content placement is performed in a decentralized manner, and the eavesdropper is unable to find any useful information about the original files. In other words, no coordination is required for the content placement. One key feature of the decentralized placement phase is that
each user's cache is filled independent of other users during the placement phase. Particularly, the placement operation of a given user neither depends on the identity nor the number of other users in the system. As a result, the users can contact different transmitters at different times for the placement phase. Having a decentralized placement phase is thus an important robustness property for a caching system\cite{Maddah_2015-decentralized}.

In this section, we generalize the secure multi-transmitter caching problem to a decentralized scenario \cite{Maddah_2015-decentralized}. Recently, a decentralized multi-transmitter coded caching scheme in linear network was investigated in \cite{WSA2021}, and \cite{Decentralized_cyclic}. In this section, we propose a method that integrates secure coded caching communication basics \cite{Secure Delivery} with a decentralized multi-transmitter scheme \cite{WSA2021}.

In secure decentralized multi-transmitter coded caching problem, each user, independent of other users, caches a fraction of bits of each file uniformly at random in the placement phase. Then, the transmitters map the end-users' cache to each files' segments. After this phase, the centralized key placement procedure is performed. During this phase, the transmitters fill each end-user's memory with some random keys \cite{one_time_pad}. The key placement phase is conducted in a centralized manner to keep safe the integrity of the keys and deter a totally passive eavesdropper from obtaining any information about the original files.

After revealing users' request, the transmitters generate the coded messages to serve multiple end-users simultaneously using the greedy decentralized algorithm presented in Algorithm \ref{alg:SDMCC}. The transmitters encode the multi-cast messages with the shared random keys and broadcast them in the network. Each legitimate client can decode the desired sub-files of the requested file, using the keys which has cached in the centralized key placement, and their cache. Algorithm \ref{alg:SDMCC} shows the pseudo-code of our scheme.

\begin{remark}
	In the case of decentralized coded caching, similar to the centralized scenario, the conventional secure scheme is one which caches only one unique key per user, and benefits from only the local caching gain provided by encrypted uni-cast delivery. This way, the coding delay can be given by $K(1-q)$.
\end{remark}

\begin{thm}\label{Theorem_Decentralized}
	For a decentralized multi-transmitter coded caching scheme, the following coding delay is securely achievable:
	\begin{align}
	 T_{S}^{D}=\sum\limits_{s=1}^{K}{\frac{\left( \begin{smallmatrix}
		 K \\ 
		 \min (s+L-1,K) \\ 
		\end{smallmatrix} \right){{(q)}^{s-1}}{{(1-q)}^{K-s+1}}}{\left( \begin{smallmatrix}
		 K-s \\ 
		 \min (s+L-1,K)-s \\ 
		\end{smallmatrix} \right)}} \times	\left( \begin{smallmatrix}
	  \min (s+L-1,K)-1 \\ 
	 s-1 \\
	\end{smallmatrix} \right),
	\label{eq_T}
	\end{align}
	such that; $M={{M}_{D}}+{{M}_{K}}$, ${{M}_{D}}=N.q$ and
	\begin{align}
	{{M}_{K}}=\sum\limits_{s=1}^{K}&{\frac{\left( \begin{smallmatrix}
			 K-1 \\ 
			 \min (s+L-1,K)-1 \\ 
			\end{smallmatrix} \right)\left( \begin{smallmatrix}
			 \min (s+L-1,K)-1 \\ 
			 s-1 \\ 
			\end{smallmatrix} \right)}{\left( \begin{smallmatrix}
			 K-s \\ 
			 \min (s+L-1,K)-s \\ 
			\end{smallmatrix} \right)}}\times{{(q)}^{s-1}}{{(1-q)}^{K-s+1}},
	\label{eq_M_Decentralized}
	\end{align}
	where ${M}_{K}$ and ${M}_{D}$ are the key and data storage size, respectively.
	In the rest of this section, we derive \eqref{eq_T} according to Algorithm \ref{alg:SDMCC}. Also, the security proof of Algorithm \ref{alg:SDMCC} and the details on how to derive \eqref{eq_M_Decentralized} are provided in Appendix D.
	
	 According to the proposed decentralized scheme in Algorithm \ref{alg:SDMCC}, each user is allowed to store any random subset of $qF$ bits of any file ${W}_{n}$. We highlight that $q$ shows the probability that each bit of file  ${W}_{n}$ will be cached at a given user. The selection of these subsets is uniform. The parameter $q$ can be easily found in a numerical manner as it is described later via an example.
\end{thm}
%\begin{proof}
%	Please refer to Appendix B for the proof of security.
%\end{proof}

We highlight that in the above Theorem, ${M}_{K}$ represents the key storage size and guarantees the system security. If one set the parameter $L$ equal to one, the equations \eqref{eq_T} and \eqref{eq_M_Decentralized} reduce to the case in \cite{Secure Delivery}. Also, \cite{WSA2021} is a special case of our work if one ignores the security guarantee.

The above theorem is defined for $M > 1$. In the case of $M=1$ (${M}_{D}=0,{M}_{K}=1$), each end-user only stores a unique key with the same size of each file to receive the messages securely in the delivery phase. Note that the secure caching scheme is impossible for $M=0$.

\begin{algorithm}[ht!]
	\scriptsize
	\caption{Secure Decentralized Multi-Transmitter Coded Caching Algorithm }\label{alg:SDMCC}
	
	\begin{algorithmic}[1]
		\STATE \textbf{Decentralized Cache Placement $({{W}_{1}},...,{{W}_{N}})$}

		\FOR{$k \in [K], n \in [N]$}
		\STATE user $k$ independently caches a subset of $q$$\hspace{0.1mm}$$F$ bits of file $ n $, chosen uniformly at random.
		\ENDFOR
		
		\STATE \textbf{Delivery Procedure for request $(d_1,... ,d_K )$ }
		\STATE \textit{Centralized Key Placement}:
		\STATE {Transmitters map the cache contents to the fragments of files $({{W}_{1}},...,{{W}_{N}})$, and generates keys as follows:}
		\FOR {$i=0,1,2,...,K$}
		\FOR {$n=1,2,...,N$}
		\STATE ${W}_{n}=\{{W}_{n,\tau};\tau \subseteq \{1,...,K\},\left| \tau  \right|=i\}$ such that ${W}_{n,\tau}$ is cached at user $k$ if $k \in \tau$.
		\ENDFOR
		\ENDFOR
		\FOR {$s=1,2,...,K$}
		\FOR {$S\subseteq \{1,2,...,K\}:\left| S \right|=\min (s+L-1,K)$}
		\STATE Key $K_{S,\left| s \right|=s}^{(\beta)};\beta =1,...,\left( \begin{smallmatrix}
			 \min (s+L-1,K)-1 \\ 
			 s-1 \\ 
			\end{smallmatrix} \right)$ is generated
		\STATE Key $K_{S,\left| s \right|=s}^{(\beta)}$ is placed in the cache of user $k$ if $k \in S$
		\ENDFOR
		\ENDFOR
		
		\STATE \textit{Coded Secure Delivery}:
		%\STATE $x \leftarrow 1 $
		\FOR{$s= K,K-1, ... ,1$}
		\STATE $ c_u  \leftarrow 1 $
		\FOR{  $ S \subseteq [ K ] ,  \lvert S \rvert = \min \{ s + L - 1 , K \}$ }
		\FOR{$\omega = 1 , 2 , ... , \binom{ \min{(s + L - 1,K)}-1 }{ s - 1 } $}
		%\STATE $ x \leftarrow x+1 $
		\FOR{$ \mathcal{U} \subseteq [ S ], \lvert \mathcal{U} \rvert = s  $}
		%\STATE $ c_{u} \leftarrow c_{u} +1 $
		%\STATE Design $ \boldsymbol{\psi} ^{ S\setminus \mathcal{U} }  $ for all $ %i \in S \setminus \mathcal{U} :  \boldsymbol{h_i} \perp \boldsymbol{\psi}^{ %S \setminus \mathcal{U} } $
		\STATE Design $\mathbf{u}_{S}^{\mathcal{U}}$ such that: for all $j\in S$, ${\mathbf{h}_{j}}\perp \mathbf{u}_{S}^{\mathcal{U}}$ if $j \notin {\mathcal{U}}$ and ${\mathbf{h}_{j}}\notperp \mathbf{u}_{S}^{\mathcal{U}}$ if $j \in {\mathcal{U}}$
		\STATE $ G_{\omega} ( \mathcal{U} ) \leftarrow \varphi^{(\omega)}_{ r \in \mathcal{U} } \left( W^{c_u}_{d_{r} , \mathcal{U} \setminus \{r\}} \right) $
		\ENDFOR
		\STATE  $ \mathbf{x}_{\omega}(S) \leftarrow \sum_{\mathcal{U} \subseteq [ S ] } ({\mathbf{u}_{S}^{{{\mathcal{U}}}}} G_{\omega} ( \mathcal{U} ))\oplus {{{K}}_{S,\left| s \right|=s}^{(\beta)}{\mathbf{w}}} $
		\ENDFOR
		\STATE transmitters send the block of:\\ $\mathbf{X}(S)=[{\mathbf{x}_{1}}(S),...,{\mathbf{x}_{\left( \begin{smallmatrix} 
					\min (s+L-1,K)-1 \\ 
					s-1 
				\end{smallmatrix} \right)}}(S)]$\\
		 $ c_{u} \leftarrow c_{u} +1 $
		\ENDFOR
		\ENDFOR
	\end{algorithmic}
\end{algorithm}
	
In the rest of this section, we present the proposed scheme's prefetching and content delivery phases and then provide the formal proof of the secure coding delay of Theorem \ref{Theorem_Decentralized}.

\textbf{Prefetching Strategy}:
During the placement phase, each user, uniformly at random and independent of other users, stores $qF$ bits of each file in the memory. Then, the transmitters map each file's fragments, which contain non-overlapping combinations of bits, with the contents of each user's cache. The fragments show which user (or group of users) has cached the particular fragment's bits. After this phase, the transmitters store secret shared keys in each user's cache as part of a centralized key placement process. To preserve the key integrity and protect the files from the eavesdropper, the key placement needs to be centrally managed. The key placement procedure is organized as follows:
\vspace{-2mm}
\begin{align}
Generate \hspace{1.5mm} K_{S,\left| s \right|=s}^{(\beta)};&S\subseteq \{1,2,...,K\}:\left| S \right|=\min (s+L-1,K),\hspace{3mm}s=1,2,...,K,\nonumber\\
&\beta =1,...,\left( \begin{smallmatrix}
\min (s+L-1,K)-1 \\ 
s-1
\end{smallmatrix} \right).
\end{align}

Key $K_{S,\left| s \right|=s}^{(\beta)}$ is placed in the cache of user $k$ if $k \in S$.

\textbf{Content Delivery Strategy}:
Consider an arbitrary $\min(s+L-1,K)$-subset of users denoted by $S$ (i.e. $S\subseteq [K],\left| S \right|=\min (s+L-1,K)$), in which $s$ takes value from $K$ to $1$, respectively. For this specific subset $S$ denote all $\mathcal{U}$ subsets of $S$ by ${\mathcal{U}_{i}}:i=1,...,\left( \begin{smallmatrix}
& \min (s+L-1,K) \\ 
& s \\ 
\end{smallmatrix} \right)$ (i.e. $\mathcal{U}_{i}\subseteq S,\left| \mathcal{U}_{i} \right|=s$). Then, for each given set of $S$ and the corresponding $\mathcal{U}_{i}$, we design zero-forcing procedure as follows:
\begin{align}
	\mathbf{u}_{S}^{{{\mathcal{U}_{i}}}}\bot \hspace{1mm} {\mathbf{h}_{j}} \hspace{1mm} for \hspace{1mm} all \hspace{1mm} j\in S\backslash {{\mathcal{U}_{i}}},\hspace{3mm}
	\mathbf{u}_{S}^{{{\mathcal{U}_{i}}}} \notperp \hspace{1mm} {\mathbf{h}_{j}} \hspace{1mm} for \hspace{1mm} all \hspace{1mm} j\in {{\mathcal{U}_{i}}},
\end{align}
while the aforementioned equations can be solved directly \cite{Shariatpanahi_2016,Multi-antenna-coded-caching}.

For each $\mathcal{U}_{i}$, we define $G(\mathcal{U}_{i})$ as an operator to extract all mini-files that each one is assigned to a subset $\mathcal{U}_{i}$ with the size of $s$:
\begin{align}
\label{formul-G}
G(\mathcal{U}_{i})={{\varphi }_{r\in {\mathcal{U}_{i}}}}(W_{{{d}_{r}},{\mathcal{U}_{i}}\backslash \{r\}}^{{{c}_{u}}}),\hspace{3mm}{{c}_{u}}=\left( \begin{smallmatrix}
 K-\left| {\mathcal{U}_{i}} \right| \\ 
 \left| S \right|-\left| {\mathcal{U}_{i}} \right|
\end{smallmatrix} \right),
\end{align}
where $W_{{{d}_{r}},{\mathcal{U}_{i}}\backslash \{r\}}^{{{c}_{u}}}$ represents a mini-file which is available in the cache of all users in $\mathcal{U}_{i}$, except $r$, and is required by user $r$.

Note that the index ${c}_{u}$ is chosen such that such mini-files have not been repeated in the former $\min(s+L-1,K)$-subsets. Also, $\varphi$ operator presents a random linear combination of the corresponding mini-files for all $r\in \mathcal{U}_{i}$. Subsequently, we define a secure coded message, assigned to a certain $\min(s+L-1,K)$-subset of users (i.e., $S$):
\begin{align}
\mathbf{x}(S)=\sum\nolimits_{\mathcal{U}\subseteq S, \left| \mathcal{U} \right|=s}(G(\mathcal{U}){\mathbf{u}_{S}^{{{\mathcal{U}}}}})\oplus K_{S,\left| s \right|=s}{\mathbf{w}},
\end{align}
where ${K}_{S}$ is the corresponding key, and ${\mathbf{w}}$ is a $L$-by-$1$ vector of ones. Note that in this scheme, at each time slot, the same keys are placed on $L$ transmitters. As a result of existing a single-antenna eavesdropper in the system, he/she cannot find any useful information about the original files. The benefit of this work is reducing the key storage by the factor of $L$. The proof of security is provided in Appendix D, subsection B.

The transmitters repeat the above procedure $\left( \begin{smallmatrix}
& \min (s+L-1,K)-1 \\ 
& s-1 \\ 
\end{smallmatrix} \right)$ times for the given $\min(s+L-1,K)$-subset $S$ so as to drive different independent versions of ${\mathbf{x}_{\omega }}(S):\omega =1,...,\left( \begin{smallmatrix}
& \min (s+L-1,K)-1 \\ 
& s-1 \\ 
\end{smallmatrix} \right).$ By way of explanation, the only difference between ${\mathbf{x}_{\omega }}(S)$'s is in the random coefficients which are chosen for calculating  the linear combinations in (\ref{formul-G}). Selecting the random coefficients results in creating independent linear combinations of the corresponding mini-files with the high probability. Therefore, to make different versions of ${\mathbf{x}_{\omega }}(S)$, we define:
\begin{align}
{\mathbf{x}_{\omega }}(S)=\sum\nolimits_{\mathcal{U}\subseteq S, \left| \mathcal{U} \right|=s}({{G}_{\omega }}(\mathcal{U}){\mathbf{u}_{S}^{{\mathcal{U}}}})\oplus K_{S,\left| s \right|=s}^{(\beta)} {\mathbf{w}},\hspace{3mm}
{{G}_{\omega }}({\mathcal{U}_{i}})=\varphi _{r\in {\mathcal{U}_{i}}}^{\omega }(W_{{{d}_{r,{\mathcal{U}_{i}}\backslash \{r\}}}}^{{{c}_{u}}}).
\end{align}
Eventually, for this specific $\min(s+L-1,K)$-subset $S$, the transmitters send the following block:
\vspace{-5mm}
\begin{align}
[{\mathbf{x}_{1}}(S),...,{\mathbf{x}_{\left( \begin{smallmatrix} 
		\min (s+L-1,K)-1 \\ 
		s-1 
		\end{smallmatrix} \right)}}(S)].
\end{align}

Index ${c}_{u}$ will be updated for those mini-files which have appeared in the linear combinations in (\ref{formul-G}). When the above procedure for this certain subset $S$ is finished, another $\min(s+L-1,K)$-subset of users is taken into account, and the same procedure is conducted for that subset and this process is repeated until all $\min(s+L-1,K)$-subsets of $[K]$ are considered.

Now, let's calculate the secure coding delay of the proposed scheme. For a fixed $\min(s+L-1,K)$-subset $S$, each $\mathbf{x}_{\omega}(S)$ is a $L$-by-$\dfrac{{{(q)}^{s-1}}{{(1-q)}^{K-s+1}}}{\left( \begin{smallmatrix}
	& K-s \\ 
	& \min (s+L-1,K)-s \\ 
	\end{smallmatrix} \right)}f$ block of symbols. Therefore, the transmit block for $S$, i.e. $[{\mathbf{x}_{1}}(S),...,{\mathbf{x}_{\left( \begin{smallmatrix} 
		\min (s+L-1,K)-1 \\ 
		s-1 
		\end{smallmatrix} \right)}}(S)]$, is a $L$-by-$\dfrac{{{(q)}^{s-1}}{{(1-q)}^{K-s+1}}}{\left( \begin{smallmatrix}
	& K-s \\ 
& \min (s+L-1,K)-s \\ 
\end{smallmatrix} \right)}\left( \begin{smallmatrix}
& \min (s+L-1,K)-1 \\ 
& s-1 \\ 
\end{smallmatrix} \right)$ block. Due to the fact that this transmission should be repeated for all $\left( \begin{smallmatrix}
& K \\ 
& \min (s+L-1,K) \\ 
\end{smallmatrix} \right)$ $\min(s+L-1,K)$-subsets of users, and for all $s$, the whole transmit block size will be:
\begin{align}
L-by-&\sum\limits_{s=1}^{K}{\frac{\left( \begin{smallmatrix}
		& K \\ 
		& \min (s+L-1,K) \\ 
		\end{smallmatrix} \right)\left( \begin{smallmatrix}
		& \min (s+L-1,K)-1 \\ 
		& s-1 \\ 
		\end{smallmatrix} \right)}{\left( \begin{smallmatrix}
		& K-s \\ 
		& \min (s+L-1,K)-s \\ 
		\end{smallmatrix} \right)}}\times{{(q)}^{s-1}}{{(1-q)}^{K-s+1}}f,
\end{align}
which will result in the following secure coding delay:
\begin{align}
T_{S}^{D}=&\sum\limits_{s=1}^{K}{\frac{\left( \begin{smallmatrix}
		& K \\ 
		& \min (s+L-1,K) \\ 
		\end{smallmatrix} \right)\left( \begin{smallmatrix}
		& \min (s+L-1,K)-1 \\ 
		& s-1 \\ 
		\end{smallmatrix} \right)}{\left( \begin{smallmatrix}
		& K-s \\ 
		& \min (s+L-1,K)-s \\ 
		\end{smallmatrix} \right)}}\times {{(q)}^{s-1}}{{(1-q)}^{K-s+1}}.
\end{align}

We now review the aforementioned scheme with an illustrative example.\\
\textbf{Example 1:} Consider a network including $K=3$ users, $L=2$ transmitters, cache size $M=2$, and a library of $N=3$ files, namely $A, B$ and $C$. In the placement phase, each user caches a subset of $qF$ bits of each file uniformly at random.
The actions of the placement procedure effectively partition any file  into $2^{3}=8$ subfiles:
\vspace{-3mm}
\begin{center}
	$A=\Big( {{A}_{\phi }},{{A}_{1}},{{A}_{2}},{{A}_{3}},{{A}_{12}},{{A}_{13}},{{A}_{23}},{{A}_{123}} \Big)$.\\
\end{center}

After mapping the contents of each user's cache with the file fragments, the transmitters produce the following keys:
\vspace{-3mm}
\begin{align}
Keys=[K_{_{123,\left| s \right|=3}},K_{_{123,\left| s \right|=2}}^{(1)},K_{_{123,\left| s \right|=2}}^{(2)},
{{K}_{_{12,\left| s \right|=1}}},{{K}_{_{13,\left| s \right|=1}}},{{K}_{_{23,\left| s \right|=1}}}].
\end{align}
%\begin{center}
%\[Keys=\begin{Bmatrix}
%K_{_{123,\left| s \right|=3}},K_{_{123,\left| s \right|=2}}^{(1)},K_{_{123,\left| s \right|=2}}^{(2)},
%{{K}_{_{12,\left| s \right|=1}}},{{K}_{_{13,\left| s \right|=1}}},{{K}_{_{23,\left| s \right|=1}}}
%\end{Bmatrix}.\]
%\end{center}
Then, put the following contents in the cache of users:
\begin{align*}
&{z}_{1}=\begin{Bmatrix}
{A}_{1}, {A}_{12}, {A}_{13}, {A}_{123},
{B}_{1}, {B}_{12},\\ {B}_{13}, {B}_{123},
{C}_{1}, {C}_{12}, {C}_{13}, {C}_{123}\\
K_{_{123,\left| s \right|=3}}, K_{_{123,\left| s \right|=2}}^{(1)}, K_{_{123,\left| s \right|=2}}^{(2)}, {{K}_{_{12,\left| s \right|=1}}}, {{K}_{_{13,\left| s \right|=1}}}
\end{Bmatrix},
&{z}_{2}=\begin{Bmatrix}
{A}_{2}, {A}_{12}, {A}_{23}, {A}_{123},
{B}_{2}, {B}_{12},\\ {B}_{23}, {B}_{123},
{C}_{2}, {C}_{12}, {C}_{23}, {C}_{123}\\
K_{_{123,\left| s \right|=3}}, K_{_{123,\left| s \right|=2}}^{(1)}, K_{_{123,\left| s \right|=2}}^{(2)}, {{K}_{_{12,\left| s \right|=1}}}, {{K}_{_{23,\left| s \right|=1}}}
\end{Bmatrix},\\
&{z}_{3}=\begin{Bmatrix}
{A}_{3}, {A}_{13}, {A}_{23}, {A}_{123},
{B}_{3}, {B}_{13}, \\{B}_{23}, {B}_{123},
{C}_{3}, {C}_{13}, {C}_{23}, {C}_{123}\\
K_{_{123,\left| s \right|=3}}, K_{_{123,\left| s \right|=2}}^{(1)}, K_{_{123,\left| s \right|=2}}^{(2)}, {{K}_{_{13,\left| s \right|=1}}}, {{K}_{_{23,\left| s \right|=1}}}
\end{Bmatrix}.
\end{align*}

Let's focus on the delivery phase of Algorithm \ref{alg:SDMCC}. Without loss of generality, suppose that the vector of users' request is ${\mathbf{d}_{k}}=\{1,2,3\}$. Then, the following secure coded messages are sent:
\vspace{-3mm}
\begin{align*}
&s=3 \hspace{2mm}, \hspace{2mm} \left| S \right|=3 \hspace{2mm} , \hspace{2mm} \left| \mathcal{U} \right|=3  \hspace{2mm} , \hspace{2mm} {S}=\{1,2,3 \} :\\
&{\mathbf{x}(\{1,2,3\})}=[ {{A}_{23}}+{{B}_{13}}+{{C}_{12}}+K_{_{123,\left| s \right|=3}}{\mathbf{w}}]\\
&s=2 \hspace{2mm} ,\hspace{2mm} \left| S \right|=3 \hspace{2mm} ,\hspace{2mm} \left| \mathcal{U} \right|=2 \hspace{2mm} ,\hspace{2mm} {S}=\{1,2,3\} :\\
{\mathbf{x}(\{1,2,3\})}=&[{{\varphi }^{1}}( {{A}_{2}}+{{B}_{1}}) \mathbf{h}_{3}^{\bot }+{{\varphi }^{1}}( {{A}_{3}}+{{C}_{1}}) \mathbf{h}_{2}^{\bot }+{{\varphi }^{1}}(  {{B}_{3}}+{{C}_{2}}) \mathbf{h}_{1}^{\bot }+K_{_{123,\left| s \right|=2}}^{(1)}{\mathbf{w}},\\ &{{\varphi }^{2}}( {{A}_{2}}+{{B}_{1}}) \mathbf{h}_{3}^{\bot }+{{\varphi }^{2}}( {{A}_{3}}+{{C}_{1}}) \mathbf{h}_{2}^{\bot }+{{\varphi }^{2}}(  {{B}_{3}}+{{C}_{2}}) \mathbf{h}_{1}^{\bot }+K_{_{123,\left| s \right|=2}}^{(2)}{\mathbf{w}}]\\
&s=1 \hspace{2mm} , \hspace{2mm} \left| S \right|=2 \hspace{2mm} , \hspace{2mm} \left| \mathcal{U} \right|=1 \hspace{2mm}  , \hspace{2mm} {S}=\{1,2\} : \\
&{\mathbf{x}(\{1,2\})}=[A_{\phi }^{(1)}\mathbf{h}_{2}^{\bot }+B_{\phi }^{(1)}\mathbf{h}_{1}^{\bot }+{{K}_{_{12,\left| s \right|=1}}}{\mathbf{w}}]\\
&s=1 \hspace{2mm} , \hspace{2mm} \left| S \right|=2 \hspace{2mm} , \hspace{2mm} \left| \mathcal{U} \right|=1 \hspace{2mm}  , \hspace{2mm} {S}=\{1,3\} :\\
&{\mathbf{x}(\{1,3\})}=[ A_{\phi }^{(2)}\mathbf{h}_{3}^{\bot }+C_{\phi }^{(1)}\mathbf{h}_{1}^{\bot }+{{K}_{_{13,\left| s \right|=1}}}{\mathbf{w}}]\\
&s=1 \hspace{2mm} , \hspace{2mm} \left| S \right|=2 \hspace{2mm} , \hspace{2mm} \left| \mathcal{U} \right|=1 \hspace{2mm}  , \hspace{2mm} {S}=\{2,3\} : \\	&{\mathbf{x}(\{2,3\})}=[ B_{\phi }^{(2)}\mathbf{h}_{3}^{\bot }+C_{\phi }^{(2)}\mathbf{h}_{2}^{\bot }+{{K}_{_{23,\left| s \right|=1}}}{\mathbf{w}}]
\end{align*}

It is easy to verify that in the above example, ${M}_{D}=1.5$, ${M}_{K}=0.5$, and ${T}_{S}^{D}=0.5625$. The parameter $q$ for this example is $0.5$.
\vspace{-5mm}
\section{Numerical Evaluations}
\vspace{-3mm}
In this section, we verify the results of our work via numerical evaluations. Clearly, due to the security constraint, combining the global/local caching gain with multiplexing gain makes the presented secure coding delays higher than the insecure achievable delays in \cite{Shariatpanahi_2016,Adding_transmitters,Feedback-bottleneck,WSA2021}, for a given value of $M$, and $N$. That is the cost paid for security in the delivery phase. According to the schemes, the worst-case delay is gained at $M=1$\footnote{Although it depends on the cache size, the presented method in Theorem 2 is defined for $M>L$. Hence, the worst-case delay for this scheme is gained at $M=L$.}, where there is no data stored in the cache of users and transmitters use the unique key of each user to transmit the users' requested file securely. Another extreme case is where $M=N$, which means all files are cached in the users' memory, and no transmission delivery is needed. In this case, ${M}_{D}=N$, ${M}_{K}=0$ and the secure achievable delay is $0$.

\begin{figure*}
	\centering
	\begin{subfigure}[b]{0.3\textwidth}
		\includegraphics[scale=.4]{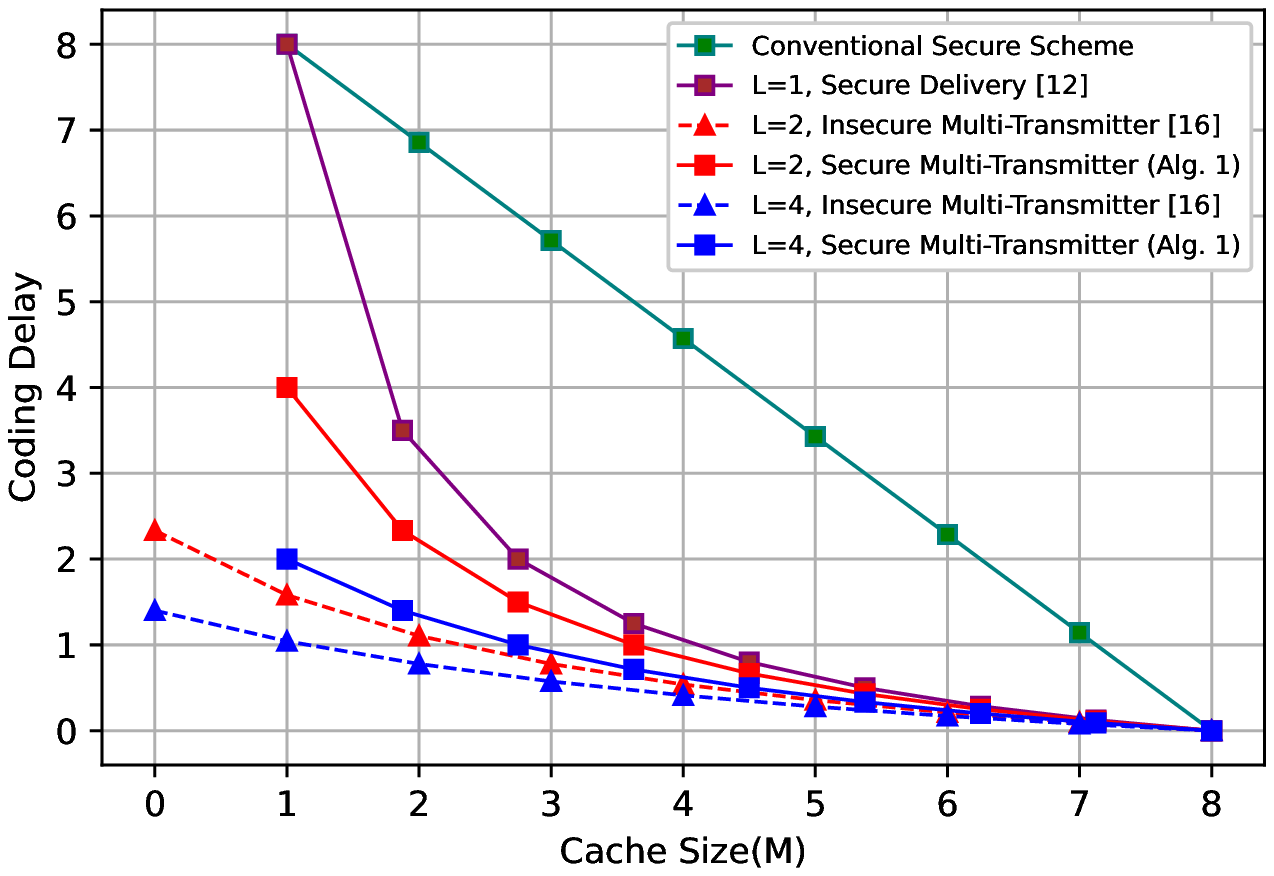}
		\caption{}
		\label{fig:figure2}
	\end{subfigure}
	\quad
	\begin{subfigure}[b]{0.3\textwidth}
		\includegraphics[scale=.4]{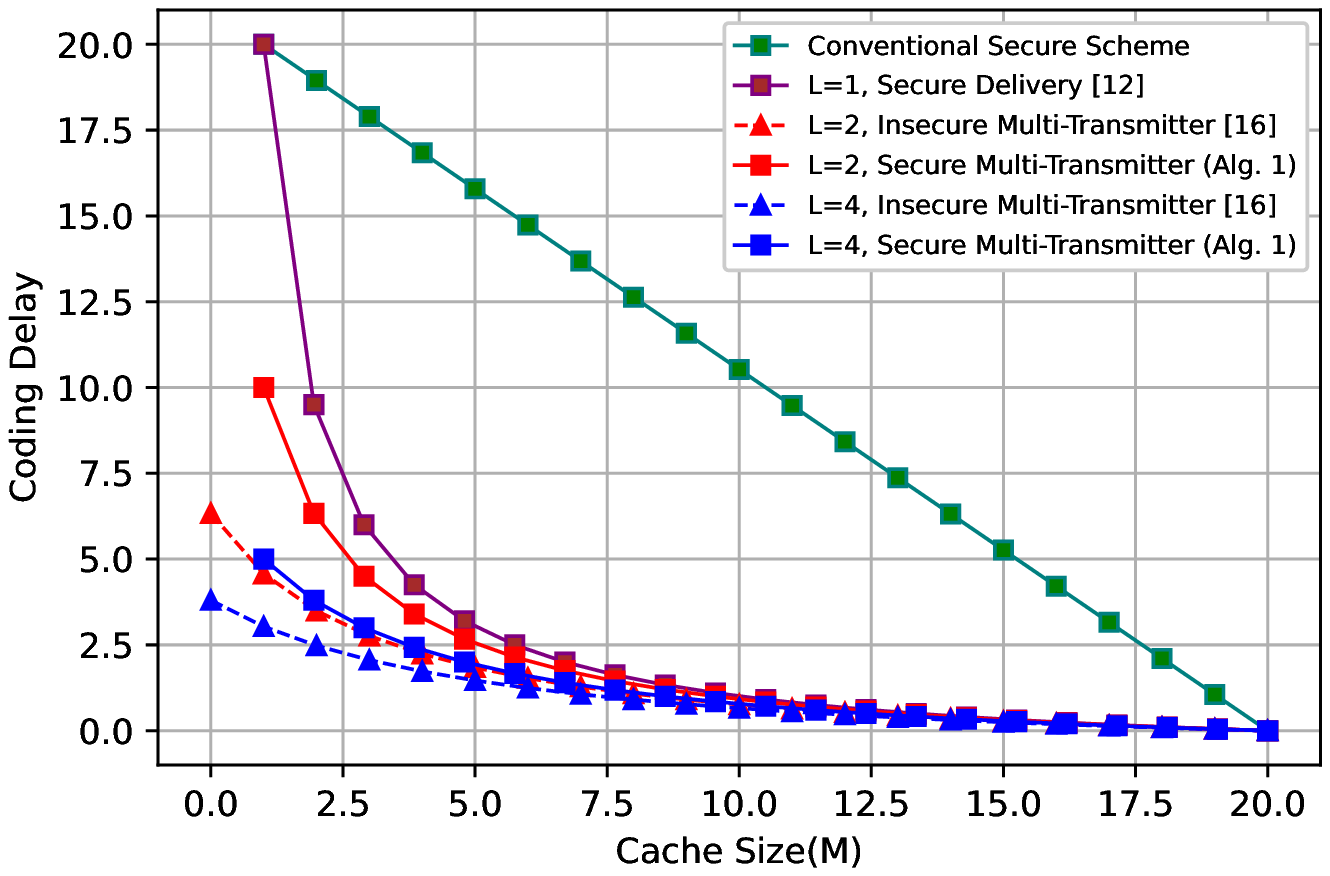}
		\caption{}
		\label{fig:figure3}
	\end{subfigure}
	\quad
	\begin{subfigure}[b]{0.3\textwidth}
		\includegraphics[scale=.4]{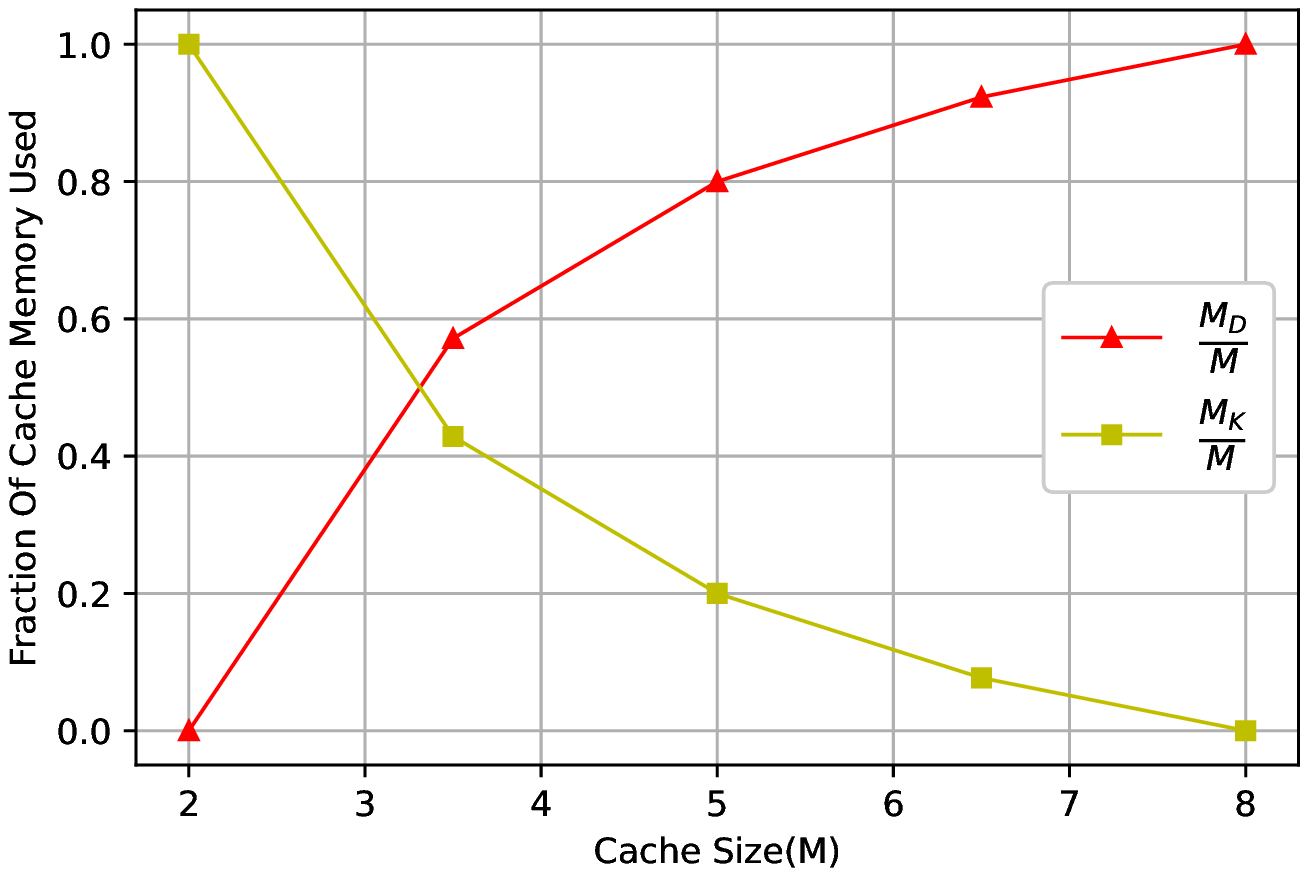}
		\caption{}
		\label{fig:figure6}
	\end{subfigure}
	\caption{(a) ($M$,${T}_{S}^{C})$ trade-off for $N=K=8$, $L=2$ and $L=4$ transmitters. Comparing secure and insecure bounds of Algorithm 1, (b) $(M$,${T}_{S}^{C})$ trade-off for $N=K=20$, $L=2$ and $L=4$ transmitters. Investigating the effect of increasing the number of files and users on secure, and insecure bounds of Algorithm 1, (c) ${M}_{K}$ vs. ${M}_{D}$ trade-off for $N=K=8$ and $L=2$ transmitters for the achievable scheme in Theorem 2.}\label{main lable}
\end{figure*}

By examining Fig. \ref{fig:figure2}, we find out that there is a gap between the secure achievable coding delay and the insecure one. Moreover, it is obvious that by increasing the number of transmitters, the secure achievable coding delay decreases and the gap between secure and insecure schemes declines. From Fig. \ref{fig:figure2}, we can see that our algorithm outperforms the scheme presented in \cite{Secure Delivery} for $L=1$ transmitter.

Comparing Fig. \ref{fig:figure2} and \ref{fig:figure3} demonstrates that the gap between secure and insecure schemes decreases when the network is scaled up. It means the security of the delivery phase is achieved with almost negligible memory cost for a large number of files, and users.

Fig. \ref{fig:figure6} shows the trade-off between the fraction of storage used by the data and keys in secure multi-transmitter coded caching system with reduced subpacketization (presented in Theorem 2) for $N=8$ files and $K=8$ users and $L=2$ transmitters. Consider the cache memory constraint in Theorem 2 i.e., $M\in \frac{(N-L)}{K}t+L$, $\forall t \in \{0,1,2,...,K\}, L|t, L|K$ jointly. Since $M={M}_{D}+{M}_{K}$, from Theorem 2, we have ${M}_{K}=L\times(1-\frac{t}{K})$, ${M}_{D}=\frac{N.t}{K}$. Fig. \ref{fig:figure6} illustrates that ${M}_{K}$ dominates at lower values of $M$. Formally, $M\ge \frac{2NL}{N+L}$, data storage dominates key storage i.e., ${M}_{D}>{M}_{K}$. In this scheme, we have $L\times \left( \begin{smallmatrix}
	\sfrac{K}{L} \\ 
	\sfrac{t}{L}+1 \\ 
\end{smallmatrix} \right)$ unique keys in the system. Thus, the case for there being only $L$ unique keys in the system corresponds to $t=K-L$ i.e., ${M}_{K}=\frac{L^2}{K}$. Hence, for avoiding $L$ shared keys across all users, we need ${M}_{K}>\frac{L^2}{K}\Rightarrow  t \le K-L$, which corresponds to $M \le \frac{(K-L)(N-L)}{K}+L$. It is also unsuitable that new keys be redistributed to the entire system, each time a user leaves. The proposed scheme avoids this case by sharing the keys between users. If a user leaves the system or is compromised, only the corresponding keys to its cache need to be replaced, leaving the others untouched. Therefore, a suitable region of operation would be:
\begin{align}
\frac{2NL}{N+L} \le M \le \frac{(K-1)(N-L)}{K}+L.
\end{align}
the above formula is reduced to the special case of our work in \cite{Secure Delivery} when the parameter $L$ is set to one. Also, it is obvious that each user has one unique key at $t = 0$, $M=L=2$. In this case, the sniffer will require access to all memories to compromise the system's security.

\begin{figure*}
	\centering
	\begin{subfigure}[b]{0.3\textwidth}
		\includegraphics[scale=.4]{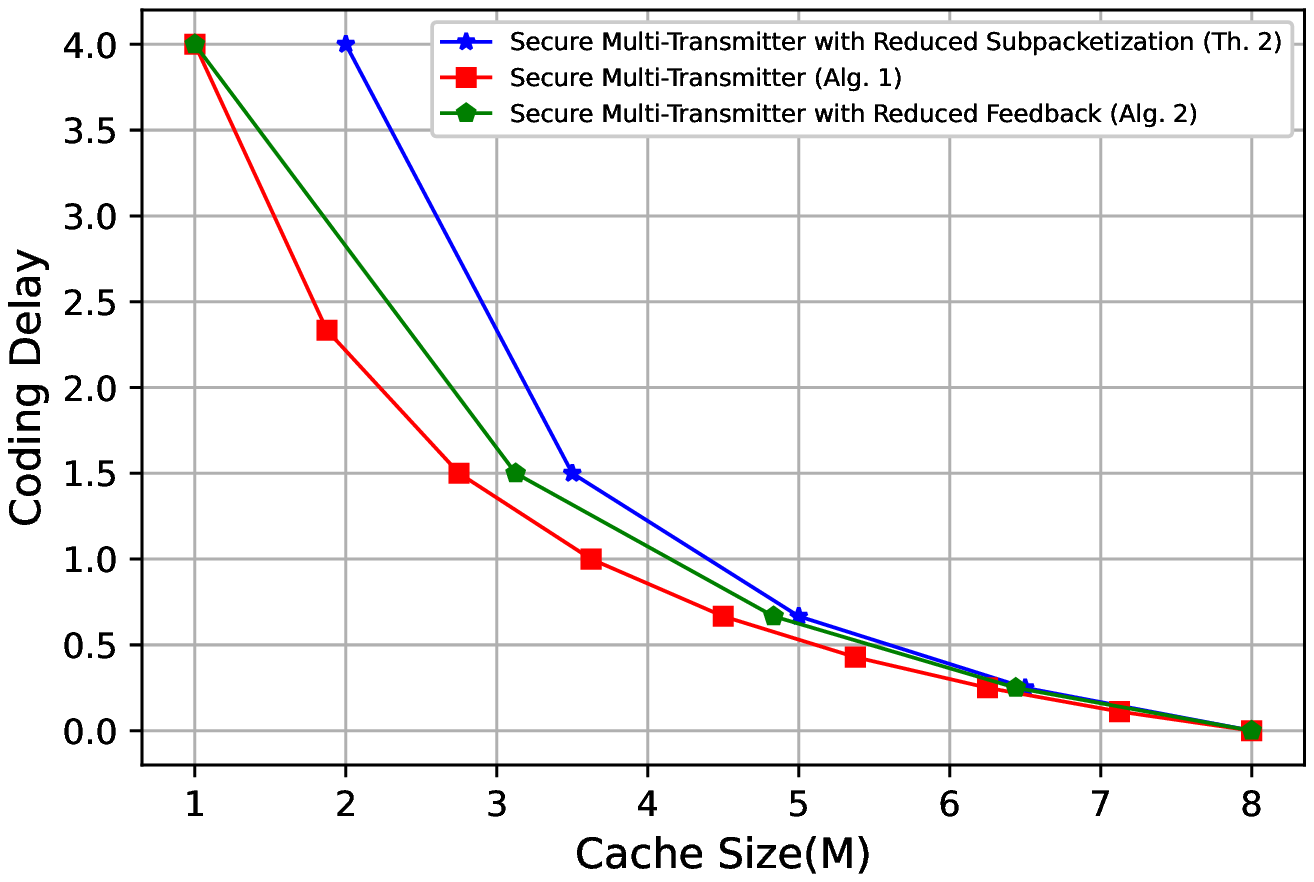}
		\caption{}
		\label{fig:figure5}
	\end{subfigure}
	\quad
	\begin{subfigure}[b]{0.3\textwidth}
		\includegraphics[scale=.4]{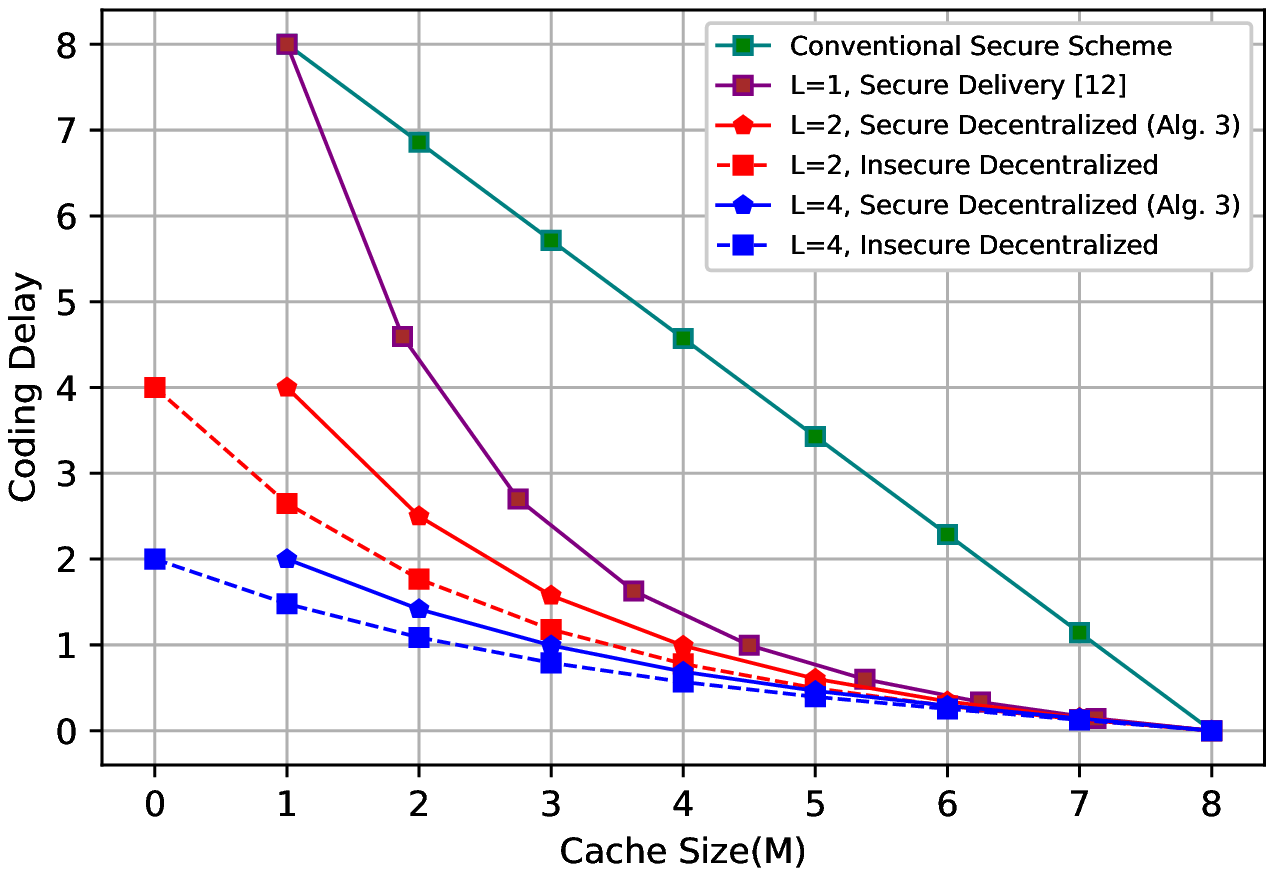}
		\caption{}
		\label{fig:figure7}
	\end{subfigure}
	\quad
	\begin{subfigure}[b]{0.3\textwidth}
		\includegraphics[scale=.4]{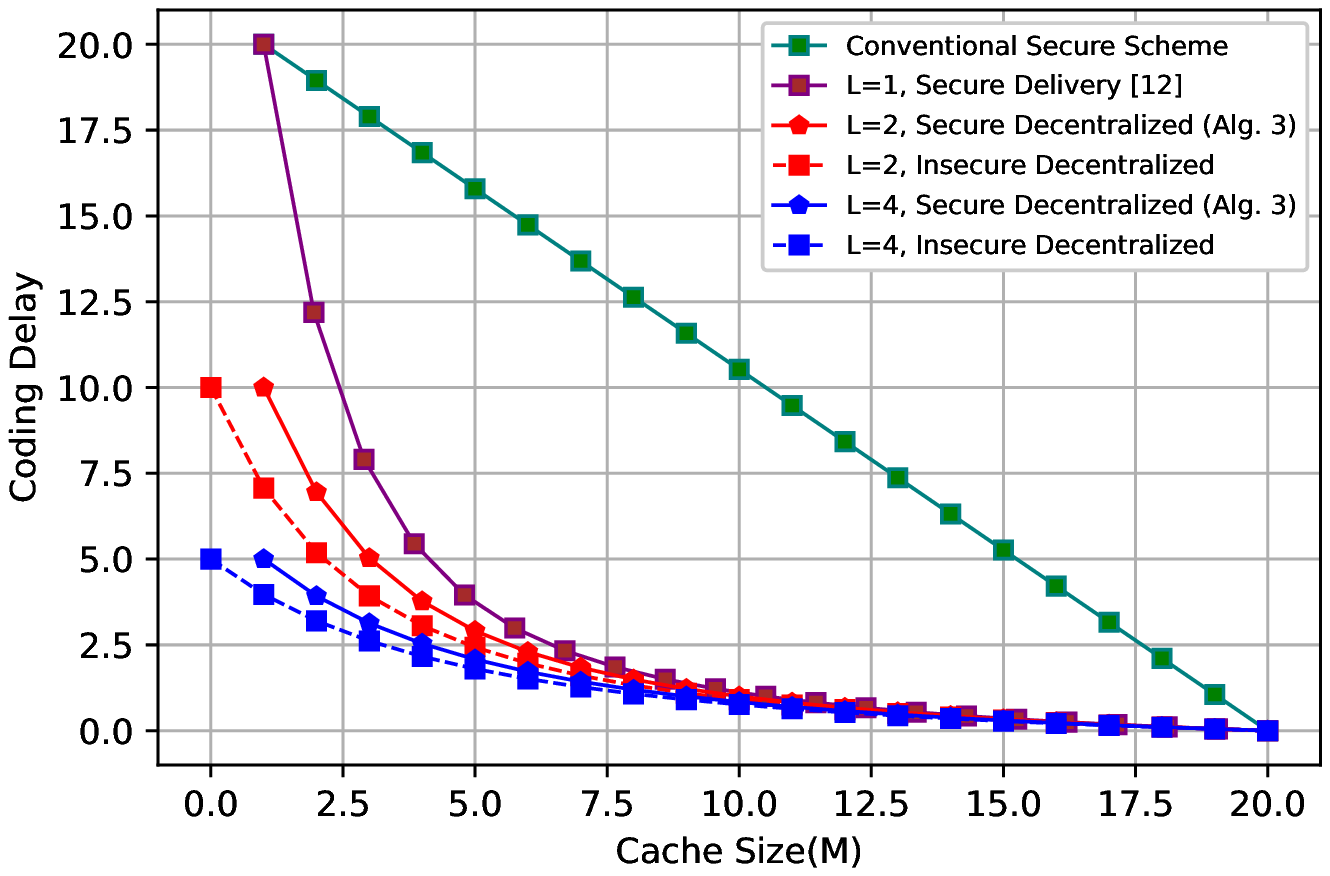}
		\caption{}
		\label{fig:figure8}
	\end{subfigure}
	\caption{(a) ($M$,${T}_{S}^{C})$ trade-off for $N=K=8$, and $L=2$ transmitters. Comparing three centralized scenarios, (b) ($M$,${T}_{S}^{D})$ trade-off for $N=K=8$, $L=2$ and $L=4$ transmitters. Comparing secure, and insecure bounds of Algorithm 3, (c) ($M$,${T}_{S}^{D})$ trade-off for $N=K=20$ and $L=2$ and $L=4$ transmitters. Investigating the effect of increasing the number of files, and users on secure and insecure bounds of Algorithm 3.}\label{main lable}
\end{figure*}

Fig. \ref{fig:figure5} shows the trade-off between secure achievable coding delays and different $M$ values for $N=K=8$ and $L=2$. By comparing three centralized proposed schemes in this paper, we can figure out that the proposed method in Algorithm 1 reaches the best performance in terms of coding delay and key storage. The proposed method in Theorem 3 has the better performance than Theorem 2 and worse than presented scheme in Algorithm 1. If ${{M}_{K}}\{L=1\}$ demonstrates the key storage of single-stream scenario, we have:
\begin{align}
&{{M}_{K}}\{Alg. 1\}=\frac{K-t}{K}={{M}_{K}}\{L=1\},\\
\vspace{3mm}
&{{M}_{K}}\{Th. 2\}=L\times (\frac{K-t}{K})=L\times {{M}_{K}}\{L=1\}= L \times {{M}_{K}}\{Alg. 1\},\\
\vspace{3mm}
&{{M}_{K}}\{Th. 3\}=\dfrac{L(K-t)(t+1)}{K(t+L)}=\dfrac{L(t+1)}{(t+L)}{{M}_{K}}\{L=1\}= \dfrac{L(t+1)}{(t+L)}{{M}_{K}}\{Alg. 1\}.
\end{align}
where ${{M}_{K}}\{Alg. 1\}$ is associated with the key storage in Algorithm 1 whereas ${{M}_{K}}\{Th. 2\}$ and ${{M}_{K}}\{Th. 3\}$ are the required key storage presented in Theorem 2 and Theorem 3, respectively.

It is remarkable that the presented method in Algorithm 1 and Theorem 3 works for $M>1$ whereas the proposed method in Theorem 2 is only defined for $M>L$. We remark that the gap between these curves decreases by increasing the number of files and users.

Fig. \ref{fig:figure7} demonstrates the trade-off between secure achievable coding delay and different $M$ values for $N=K=8$ and $L=2$ and $L=4$ transmitters, comparing the secure achievable coding delay of the decentralized setup in Algorithm 3 and the insecure scheme \cite{WSA2021}. It is obvious that by increasing the number of transmitters, the secure achievable coding delay decreases and the gap between secure and insecure schemes declines. From Fig. \ref{fig:figure7}, we can see that our algorithm outperform than the secure achievable scheme presented in \cite{Secure Delivery} for $L=1$ transmitter. We can see that there is a gap between the secure and insecure schemes. Compared to Fig. \ref{fig:figure8}, one can infer that this gap decreases when the number of files and users increases. This also implies that in the decentralized scheme, similar to centralized scheme, the security cost is \emph{almost negligible} when the network is scaled up.

\begin{figure*}
	\centering
	\begin{subfigure}[b]{0.3\textwidth}
		\includegraphics[scale=.4]{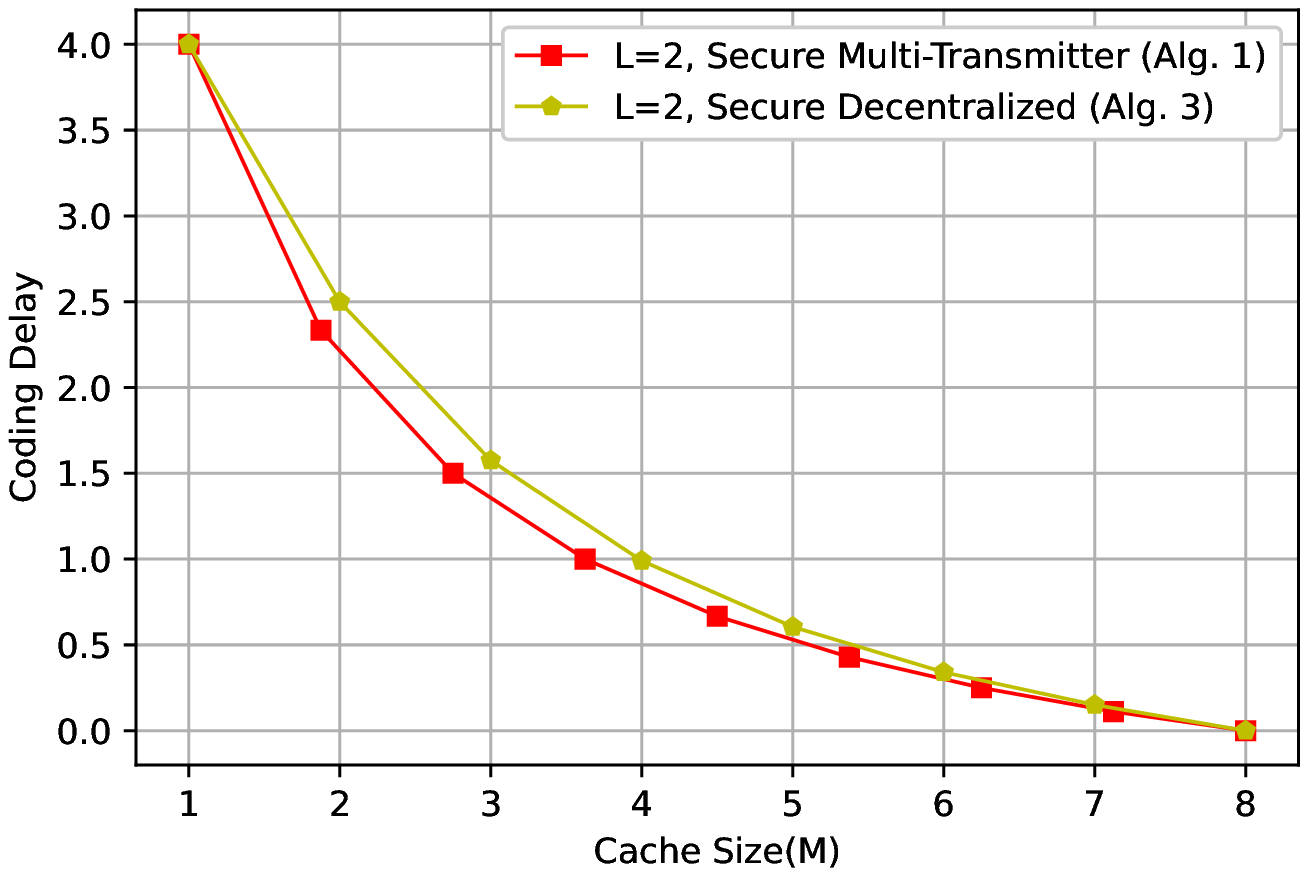}
		\caption{}
		\label{fig:figure_centralized_decentralized_1}
	\end{subfigure}
	\quad
	\begin{subfigure}[b]{0.3\textwidth}
		\includegraphics[scale=.4]{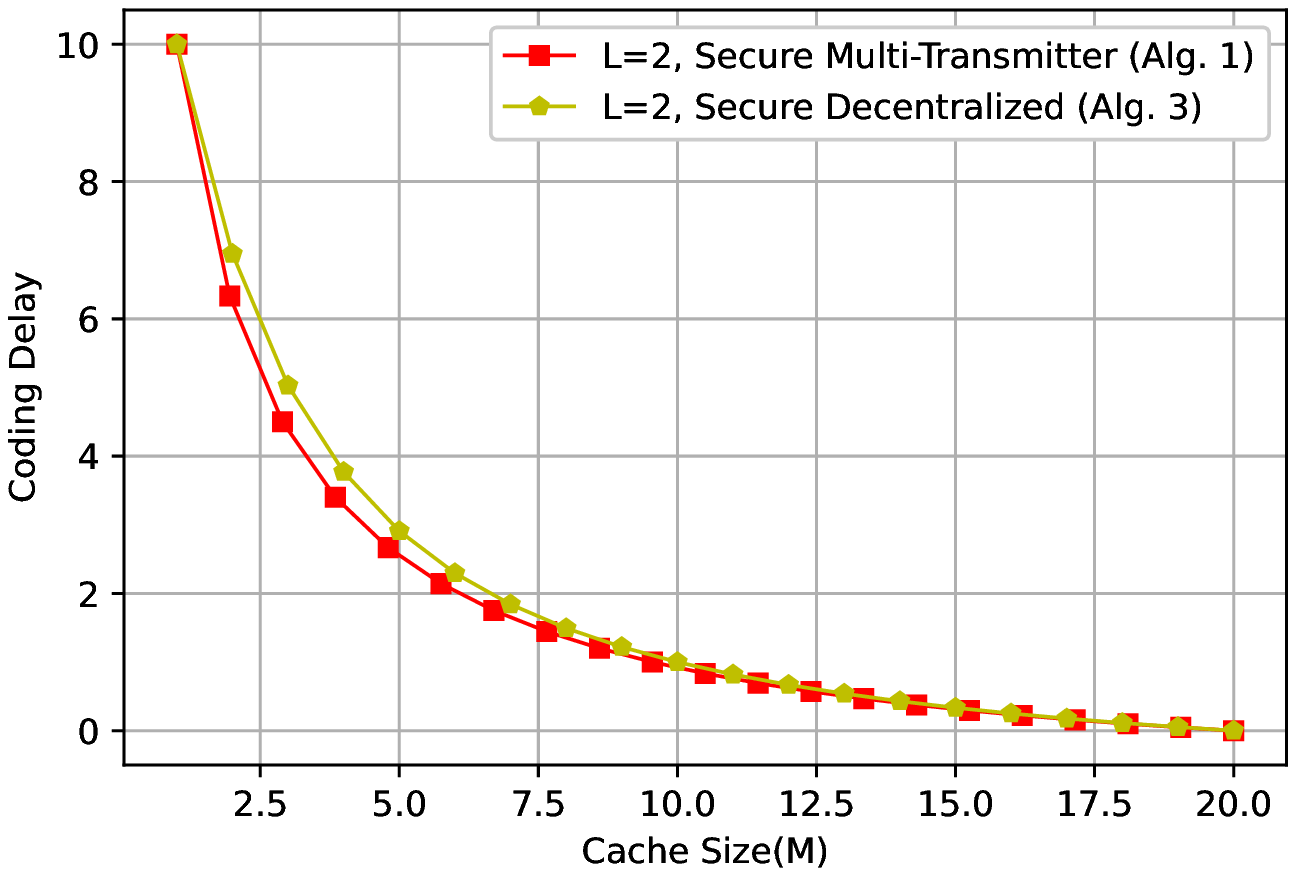}
		\caption{}
		\label{fig:figure_centralized_decentralized}
	\end{subfigure}
	\quad
	\begin{subfigure}[b]{0.3\textwidth}
		\includegraphics[scale=.4]{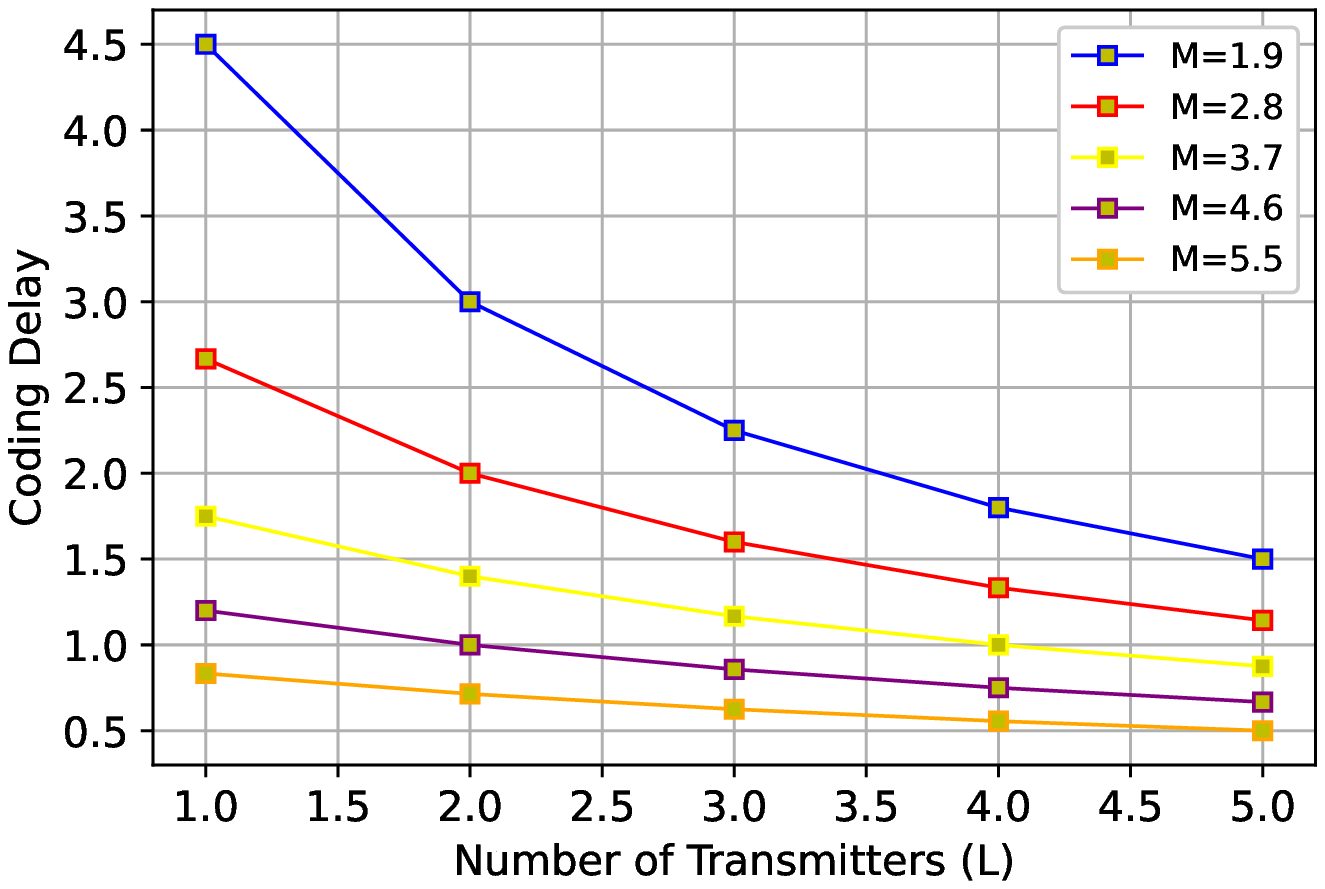}
		\caption{}
		\label{fig:figure_investigating_M}
	\end{subfigure}
	\caption{(a) Centralized vs Decentralized Secure bounds for $N=K=8$, and $L=2$ transmitters, (b) Centralized vs Decentralized Secure bounds for $N=K=20$, and $L=2$ transmitters, (c) Investigating the effect of adding transmitters on the performance of Algorithm 1, including $N=10$ files, and $K=10$ users.}\label{main lable}
\end{figure*}

Fig. \ref{fig:figure_centralized_decentralized} depicts the secure centralized and decentralized trade-off for a large number of files and users ($N=K=30$). Compared to Fig. \ref{fig:figure_centralized_decentralized_1} ($N=K=8$), we can comprehend that when the number of files and users increases, the secure decentralized scheme approaches the secure centralized scheme, presented in Algorithm 1. Hence, for a vast number of files and users, the secure coding delays are \textit{asymptotically} equal.

Fig. \ref{fig:figure_investigating_M} shows the effect of the number of transmitters on the performance of Algorithm 1, for different values of $M$. This figure clearly shows the benefits of the multiplexing gain provided by multiple transmitters in the centralized setting i.e., (increasing DoF from $t+1$ in single-stream scenario to $t+L$ in multi-transmitter setup and appearing the factor of $L$ in the denominator of secure achievable coding delay). Also, this figure shows that by increasing the memory size of each user, the secure coding delay decreases because fewer transmissions are required to send to fulfill the users' demand.

\vspace{-4mm}
\section{Conclusion AND FUTURE WORK}
\vspace{-3mm}
In this paper, we studied the secure multi-transmitter coded caching problem in the presence of a totally passive eavesdropper for both centralized and decentralized settings. We proposed key-based secure caching strategies which are robust to compromising of users and keys. The presented methods achieved the same Degrees-of-Freedom as known multi-transmitter coded caching strategies, but with secure delivery at the cost of more storage. We studied the performance of the schemes, particularly for the centralized and decentralized multi-transmitter scenarios, by using coding delay and mutual information metrics. Also, we compared the three centralized multi-transmitter scenarios in terms of key storage and coding delay. Numerical evaluations showed that for large number of files and users, the secure bounds approach the non-secure cases for both centralized and decentralized schemes i.e., the cost of security in the system is negligible when the number of files and users increases. Moreover, it has been shown that for a vast number of files and users, the secure coding delay of centralized and decentralized schemes are asymptotically equal. In our future work, we will generalize this scenario to the case of multiple randomly-located eavesdroppers \cite{multi-eav}. Also, we aim to work on the converse of the presented schemes to investigate the optimality of our proposed schemes.
\vspace{-4mm}
\appendices
\section{Corresponding Proofs of Secure Multi-transmitter Coded Caching Scheme}
\vspace{-2mm}
\subsection{Proof of Key and Data Storage}
\vspace{-2mm}
In this subsection, we discuss the data and key storage of Theorem 1. We suppose a cache size $M\le N$, and $M=\frac{N-1}{K}.\{0,...,K\}+1$. Let $t \in \{0,...,K\}$ be an integer between $0$ and $K$. The cache memory size of each user can be parameterized by $t$ as:
\vspace{-2mm}
\begin{align}
M=\frac{Nt}{K}+(1-\frac{t}{K}).
\label{eq_M_Multi_server}
\end{align}

From equation \eqref{eq_M_Multi_server}, we have $t=\frac{K(M-1)}{N-1}$. Then, the cache memory of users can be broken up into two parts, data and key storage as follows:
\vspace{-2mm}
\begin{align}
{M}_{K}=(1-\frac{t}{K}), {M}_{D}=M-{M}_{K}=\frac{Nt}{K}.
\end{align}

In the placement phase, each file $W_n$, for $n=1,...,N$, is broken up into  $\left(\begin{smallmatrix}
	K \\ 
	t \\ 
\end{smallmatrix}\right)$ non-overlapping equal-sized sub-files, according to \eqref{eq_W_Multi_server}. The sub-file $W_{n,\tau }$, for each $n$, is placed in the cache of user $k$ if $k \in \tau$. Since $\left| \tau  \right|=t$, for each user $k \in \tau$, there are $\left(\begin{smallmatrix}
  	 K-1 \\ 
	 t-1 \\ 
\end{smallmatrix} \right)$ possible users with whom it shares a sub-file of a given file $W_n$. Hence, each user stores $N\left(\begin{smallmatrix}
    K-1\\ 
	t-1 \\ 
\end{smallmatrix}\right)$ sub-files. We further divide each sub-file $W_{n,\tau }$ into $\left( \begin{smallmatrix}
K-t-1 \\ 
L-1 \\ 
\end{smallmatrix} \right)$ non-overlapping equal-sized mini-files, according to \eqref{eq_mini_file}.

 Then, the transmitters generate keys ${{K}_{{{\tau }_{k}}}^{(\beta)}}$ with the same size of mini-files, according to equation \eqref{eq_key_generation_multi_server}, uniformly at random. The key ${{K}_{{{\tau }_{k}}}^{(\beta)}}, for \hspace{1.5mm} \beta=1,..,,\left( \begin{smallmatrix}
 t+L-1 \\ 
 t \\ 
 \end{smallmatrix} \right)$, is stored in the cache of user $k$ if $k \in {\tau}_{k}$. Since  $\left| {\tau}_k \right|=t+L$ and $\left| \beta \right|=\left( \begin{smallmatrix}
 t+L-1 \\ 
 t \\ 
 \end{smallmatrix} \right)$, each user $k$ shares key ${{K}_{{{\tau }_{k}}}^{(\beta)}}$ with $\left( \begin{smallmatrix}
 K-1 \\ 
 t+L-1 \\ 
 \end{smallmatrix} \right)\left( \begin{smallmatrix}
 t+L-1 \\ 
 t \\ 
 \end{smallmatrix} \right)$ possible users. Thus, there are $\left( \begin{smallmatrix}
 K-1 \\ 
 t+L-1 \\ 
 \end{smallmatrix} \right)\left( \begin{smallmatrix}
 t+L-1 \\ 
 t \\ 
 \end{smallmatrix} \right)$ independent keys in the cache of each user. Given each key and mini-file consist of $\frac{f}{\left(\begin{smallmatrix}
 	K \\ 
 	t \\ 
 	\end{smallmatrix}\right)\left(\begin{smallmatrix}
 	K-t-1 \\ 
 	L-1 \\ 
 	\end{smallmatrix}\right)}$ symbols, the number of required bits for storing at each user can be calculated as:
 \vspace{-1mm}
 \begin{align}
 & N \left(\begin{smallmatrix}
 K-1 \\ 
 t-1 \\ 
 \end{smallmatrix}\right).\frac{f}{\left( \begin{smallmatrix}
 	K\\ 
 	t \\ 
 	\end{smallmatrix}\right)}+ \left(\begin{smallmatrix}
 K-1 \\ 
 t+L-1 \\ 
 \end{smallmatrix}\right) \left(\begin{smallmatrix}
 t+L-1 \\ 
 t \\ 
 \end{smallmatrix}\right).\frac{f}{\left(\begin{smallmatrix}
 	K \\ 
 	t \\ 
 	\end{smallmatrix}\right) \left(\begin{smallmatrix}
 	K-t-1 \\ 
 	L-1 \\ 
 	\end{smallmatrix}\right)}=(\frac{Nt}{K}+1-\frac{t}{K})f=Mf,
 \end{align}
 which satisfies the memory constraint.
\vspace{-5mm}
\subsection{Secure Coding Delay Calculation}
\vspace{-3mm}
In this subsection, we discuss the secure achievable coding delay of Theorem 1. For a certain ($t+L$)-subset $S$, each ${\mathbf{x}_{\omega }}(S)$ presents a $L$-by-$\frac{f}{\left(\begin{smallmatrix}
	K \\ 
	t \\ 
	\end{smallmatrix} \right)\left(\begin{smallmatrix}
	K-t-1 \\ 
	L-1 \\ 
	\end{smallmatrix}\right)}
$ block of symbols. Therefore, the transmitted block corresponding to this $S$, i.e. $[{\mathbf{x}_{1}}(S),...,{\mathbf{x}_{\left( \begin{smallmatrix} 
		t+L-1 \\ 
		t 
		\end{smallmatrix} \right)}}(S)]$, is a $L$-by-$\frac{f}{\left( \begin{smallmatrix}
	K \\ 
	t \\ 
	\end{smallmatrix} \right)\left( \begin{smallmatrix}
	K-t-1 \\ 
	L-1 \\ 
	\end{smallmatrix} \right)}\left( \begin{smallmatrix}
t+L-1 \\ 
t \\ 
\end{smallmatrix} \right)$ block. Since we should carry out the same procedure for all $\left( \begin{smallmatrix}
K \\ 
t+L \\ 
\end{smallmatrix} \right)$ ($t+L$)-subset of users, the entire transmitted block has the size of:
\vspace{-3mm}
\begin{align}
L-by-\frac{\left( \begin{smallmatrix}
	t+L-1 \\ 
	t \\ 
	\end{smallmatrix} \right)}{\left( \begin{smallmatrix}
	K \\ 
	t \\ 
	\end{smallmatrix} \right)\left( \begin{smallmatrix}
	K-t-1 \\ 
	L-1 \\ 
	\end{smallmatrix} \right)}\left( \begin{smallmatrix}
K \\ 
t+L \\ 
\end{smallmatrix} \right)f=L-by-{\large \dfrac{K(1-\frac{M-1}{N-1})}{L+K(\frac{M-1}{N-1})}f},
\end{align}
\vspace{-1mm}
which results in the following secure coding delay:
\begin{align}
{T}_{S}^{C}=\frac{K(1-\frac{M-1}{N-1})}{L+K(\frac{M-1}{N-1})}.
\end{align}

\section{Proof of Theorem 2}
In this subsection, we discuss the secure achievable coding delay of Theorem 2. Also, we derive the data and key storage size, as we stated in Theorem 2. We consider a cache size $M\le N$ and $M=\frac{N-L}{K}.\{0,...,K\}+L$, such that $L|t$ and $L|K$ jointly. Let $t \in \{0,...,K\}$ be an integer between $0$ and $K$. The cache memory size of each user can be parameterized by $t$ as:
\begin{align}
M=\frac{Nt}{K}+L(1-\frac{t}{K}).
\label{eq_M_Adding_Trans}
\end{align}

From equation \eqref{eq_M_Adding_Trans}, we have $t=\frac{K(M-L)}{N-L}$. Then, the cache memory of users can be separated into two parts, data and key storage as follows :
\begin{align}
{M}_{K}=L(1-\frac{t}{K}), {M}_{D}=M-{M}_{K}=\frac{Nt}{K}.
\end{align}

\underline{\textbf{Prefetching Strategy:}}
In this phase, each file $W_n$, for $n=1,...,N$, is segmented into  $\left(\begin{smallmatrix}
\frac{K}{L} \\ 
\frac{t}{L} \\ 
\end{smallmatrix}\right)$ non-overlapping equal-sized sub-files, according to equation \eqref{eq_fraction}. The sub-file $W_{n,\tau }$, for each $n$, is placed in the cache of user $k$ if $k \in \tau$. Since $\left| \tau  \right|=\frac{t}{L}$, for each user $k \in \tau$, there are $\left(\begin{smallmatrix}
\frac{K}{L}-1 \\ 
\frac{t}{L}-1 \\ 
\end{smallmatrix}\right) $ possible users with whom it shares a sub-file of a given file $W_n$. Hence, each user stores $N\left(\begin{smallmatrix}
\frac{K}{L}-1 \\ 
\frac{t}{L}-1 \\ 
\end{smallmatrix}\right)$ sub-files. Then, the transmitters generate keys ${{K}_{{{\tau }_{k}}}^{(\beta)}}$ with the same size of sub-files, according to equation \eqref{key_adding_trans}, uniformly at random. The key ${{K}_{{{\tau }_{k}}}^{(\beta)}}, for \hspace{1.5mm} \beta=1,..,L$, is cached in the memory of user $k$ if $k$ be a member of groups-set ${\tau}_{k}$. Since  $\left| {\tau}_k \right|=\frac{t}{L}+1$ and as a result of putting $L$ different independent keys on $L$ transmitters, each user $k$ shares key ${{K}_{{{\tau }_{k}}}^{(\beta)}}$ with $L\left( \begin{smallmatrix}
\frac{K}{L}-1 \\ 
\frac{t}{L} \\ 
\end{smallmatrix} \right)$ possible users. Thus, there are $L\left( \begin{smallmatrix}
\frac{K}{L}-1 \\ 
\frac{t}{L} \\ 
\end{smallmatrix} \right)$ keys in the cache of each user. Since each key and sub-file consist of $\frac{f}{\left( \begin{smallmatrix}
	\frac{K}{L} \\ 
	\frac{t}{L} \\ 
	\end{smallmatrix} \right)}$ symbols, the number of required bits for storing at each user can be derived as:
\begin{align}
& N \left(\begin{smallmatrix}
\frac{K}{L}-1 \\ 
\frac{t}{L}-1 \\ 
\end{smallmatrix}\right).\frac{f}{\left( \begin{smallmatrix}
	\frac{K}{L}\\ 
	\frac{t}{L} \\ 
	\end{smallmatrix}\right) }+L \left(\begin{smallmatrix}
\frac{K}{L}-1 \\ 
\frac{t}{L} \\ 
\end{smallmatrix}\right).\frac{f}{\left( \begin{smallmatrix}
	\frac{K}{L} \\ 
	\frac{t}{L} \\ 
	\end{smallmatrix}\right)}=\frac{Nt}{K}f+L(1-\frac{t}{K})f=(\frac{Nt}{K}+L(1-\frac{t}{K}))f=Mf,
\end{align}
which satisfies the memory constraint.

\underline{\textbf{Delivery Strategy:}}
This phase starts after revealing the demand vector $\mathbf{d}$. To fulfill the demands, transmitters generate the coded messages to serve $t+L$ users ($\frac{t}{L}+1$ groups), at each time slot. Let $S\subseteq [K']$ be a subset of set $K'$ with the size of  $\left| S \right|=\frac{t}{L}+1$. For a certain $(\frac{t}{L}+1)$-subset $S$, $\mathbf{x}(S)$ presents a coded message consists of $\frac{f}{\left( \begin{smallmatrix}
	\frac{K}{L} \\ 
	\frac{t}{L} \\ 
	\end{smallmatrix} \right)}$ symbols. Since we should transmit coded messages associate with all $(\frac{t}{L}+1)$-subsets of $K'$ to satisfy the demands, the number of bits sent over the linear network is calculated as:
\begin{align}
{T}_{S}^{C}=\frac{\left(\begin{smallmatrix}
	\frac{K}{L} \\ 
	\frac{t}{L}+1 \\ 
	\end{smallmatrix}\right)}{\left(\begin{smallmatrix}
	\frac{K}{L} \\ 
	\frac{t}{L} \\ 
	\end{smallmatrix}\right)}f=\frac{K(1-\frac{M-L}{N-L})}{L+\frac{K(M-L)}{N-L}}.
\end{align}

\section{Proof of Theorem 3}
In this subsection, we discuss the achievable coding delay of Theorem \ref{Theorem_Performance}, which was presented in Algorithm \ref{alg:SMACC}. Also, we derive the data and key storage size, as we stated in Theorem 3. We assume the cache size $M\le N$. The cache of users can be divided into data and key storage and characterized as follow as:
\begin{align}
{{M}_{D}}=\frac{N.t}{K}, {{M}_{K}}=\frac{L(K-t)(t+1)}{K(t+L)} ; M={{M}_{D}}+{{M}_{K}}.
\end{align}
\underline{\textbf{Prefetching Strategy:}}
In this phase, each file ${{W}_{n}}$ is fragmented into $K \choose t$ non-overlapping equal-sized sub-files, according to equation \eqref{eq_fragments_files}. The sub-file $W_{n,\tau }$ is placed in the cache of user $k$ if $k\in \tau$. Since $\left| \tau  \right|=t$, for each user $k\in \tau$, there are $K-1 \choose t-1$ possible users with whom it shares a sub-file of a given file ${{W}_{n}}$. Therefore, each user stores $N$$K-1 \choose t-1$ sub-files. Furthermore, each sub-file splits twice, according to equations \eqref{eq_key_split1} and \eqref{eq_key_split2} .Then, the transmitters generate ${{K}_{{{\tau }_{k}}}^{(\beta)}}$ with the same size of mini-files, according to equation \eqref{eq_keys_generation_feedback} .
The key ${{K}_{{{\tau }_{k}}}^{(\beta)}}$, $for \hspace{1mm} \beta=1,...,\left( \begin{smallmatrix}
K-t-1 \\ 
L-1 \\ 
\end{smallmatrix} \right)(t+1)L$, is stored in the memory of user $k$ if $k\in {\tau}_{k}$ . All keys are produced uniformly at random.

We remark that ${ \lambda (i) \cup \pi}$ in Algorithm \ref{alg:SMACC} (line 22) leads to producing the keys. To count the number of keys, we use the set $\lambda$ as a unique identifier because each key has a non-replicated corresponding $\lambda$. The size of set $\lambda$ is $L$, and each $\lambda$ related to a certain key, takes one of the members from the possible users indexed in the key, i.e., $t+1$ possible users, and the rest of the members are selected from the users who are not present in the specific key, i.e., $K-(t+1)$. Since $\left| {{\tau }_{k}} \right|=t+1$, and as a result of placing $L$ different independent secret shared keys on $L$ transmitters, each user $k\in {{\tau }_{k}}$ shares key ${{K}_{{{\tau }_{k}}}}$ with ${ \left(\begin{smallmatrix}
	K-1 \\ 
	t \\ 
	\end{smallmatrix}\right)  \left(\begin{smallmatrix}
	K-t-1 \\ 
	L-1 \\ 
	\end{smallmatrix}\right)(t+1)L}$ possible users. So, key storage size can be derived as:
\vspace{-2.1mm}
\begin{align}
{{M}_{K}}=\frac{\left(\begin{smallmatrix}
	K-1 \\ 
	t \\ 
	\end{smallmatrix}\right) \left(\begin{smallmatrix}
	K-t-1 \\ 
	L-1 \\ 
	\end{smallmatrix}\right)(t+1)L}{{S}_{L}}f,
\end{align}
where ${S}_{L}$ refers to the subpacketization procedure, and we will explain it later carefully.

\underline{\textbf{Delivery Strategy:}}
This phase starts after revealing the demand vector $\mathbf{d}$. To satisfy the demands, transmitters generate the coded messages to serve $t+L$ users. Transmitters choose a subset of $t+L$ users, and breaks it into two disjoint sets, $\lambda$ and $\pi$, according to Algorithm \ref{alg:SMACC}.

According to the subpacketization of files, each key must have the size of $\frac{f}{{S}_L}$ symbols. Thus, we can define more formally ${S}_{L}$ as a subpacketization procedure as follows:
\vspace{-2mm}
\begin{align}
{{S}_{L}}= \left(\begin{smallmatrix}
K \\ 
t \\ 
\end{smallmatrix}\right) \left(\begin{smallmatrix}
K-t-1 \\ 
L-1 \\ 
\end{smallmatrix}\right)(L+t).
\end{align}

Finally, we can define the coding delay in Theorem \ref{Theorem_Performance} as follows:
\begin{align}
{{T}_{S}}=\frac{\overbrace{ \left(\begin{smallmatrix}
		K \\ 
		L \\ 
		\end{smallmatrix}\right) }^{Set \hspace{0.5mm}\lambda }\overbrace{ \left(\begin{smallmatrix}
		K-L \\ 
		t \\ 
		\end{smallmatrix}\right) }^{Set \hspace{0.5mm}\pi }\overbrace{L}^{\hspace{1mm}L \hspace{0.75mm}iterations}}{\underbrace{ \left(\begin{smallmatrix}
		K \\ 
		t \\ 
		\end{smallmatrix}\right)  \left(\begin{smallmatrix}
		K-t-1 \\ 
		L-1 \\ 
		\end{smallmatrix}\right) (L+t)}_{{{S}_{L}}}}f=\frac{K-t}{L+t}.
\end{align}
The proof of Theorem \ref{Theorem_Performance} was inspired by the special case of our scheme proved in \cite{Feedback-bottleneck}, while we derived the closed-form expressions of ${M}_{K}$, for multi-transmitter caching strategy.
\vspace{-1mm}
\section{Corresponding Proofs of Secure Decentralized Multi-transmitter Coded Caching Scheme}
\vspace{-2mm}
\subsection{Proof of Key and Data Storage}
In this subsection, we discuss the data and key storage of Theorem 4. According to the decentralized scheme, presented in Algorithm \ref{alg:SDMCC}, each user can store any random subset of $qF$ bits of any file ${W}_{n}$. Consider a particular bit of file ${W}_{n}$. Since the selection of these subsets is uniform, by symmetry, the probability that this bit being cached at a fixed user is $q$. Thus, for a fixed subset of $s$ out of $K$ users, the probability that this bit is cached exactly at these $s$ users, and not stored at the remaining $K-s$ users is given by:
\begin{align}
q^s (1-q)^{K-s}.
\label{eq_size_q}
\end{align}

Therefore, the expected number of bits of file ${W}_{n}$ that are cached at exactly these $s$ users is:
\begin{align}
F q^s (1-q)^{K-s}.
\label{eq_size_q_F}
\end{align}

Moreover, for $F$ large enough, the actual realization of the random number of bits of the file ${W}_{n}$ cached at $s$ users is within the range:
\begin{align}
F q^s (1-q)^{K-s}\pm o(F),
\label{eq_size_q_F_O}
\end{align}
with the high probability. For the sake of presentation clarity, we consider all the fragments of files, shared by $s$ users, have the same size. So, the term $o(F)$ can
be ignored for large enough $F$.

Referring to Algorithm \ref{alg:SDMCC}, line 8, the variable $i$ conveys the number of users which share a certain file fragment. For $i=0$, the file fragments are ${W}_{n,\phi}$, for $n=1,...,N$, which are not cached at any users. For $i=1$, the file fragments are ${W}_{n,k}$, for $k=1,...,K$, which are cached only at one user, and thus shared by none. Generally, for any $i$, the file fragments ${W}_{n,S},\left| S \right|=i$ are cached at $i$ users, and shared by any given user with $i-1$ other users. Therefore, for a given user $k$, the number of file fragments that it shares with $i-1$ out of the remaining $K-1$ users for each $i$ is given by $\left( \begin{smallmatrix}
K-1 \\ 
i-1 \\ 
\end{smallmatrix} \right)$. From equation \eqref{eq_size_q_F}, the file fragments which are stored at exactly $i$ users, have the size of $F q^i (1-q)^{K-i}$. So, the total required memory at each user for caching files is given by:
\begin{align}
{{M}_{D}}F&= N.\sum\limits_{i=1}^{K}{\left(\begin{smallmatrix}
	K-1 \\ 
	i-1 \\ 
	\end{smallmatrix}\right)}F{{q}^{i}}{{(1-q)}^{K-i}}=Nq\sum\limits_{i-1=0}^{K-1}{\left(\begin{smallmatrix}
	K-1 \\ 
	i-1 \\ 
	\end{smallmatrix}\right)}{{q}^{i-1}}{{(1-q)}^{(K-1)-(i-1)}}=Nq. \label{eq_after_bionomial}
\end{align}

Next, we drive the key storage size which was mentioned in equation \eqref{eq_M_Decentralized}. As we said before, key $K_{S,\left| s \right|=s}^{(\beta)}$ is placed in the cache of user $k$ if  $k \in S$. Since, the key $K_{S,\left| s \right|=s}^{(\beta)}$ is associate with the transmission of ${\mathbf{x}_{\omega }}(S)=\sum\nolimits_{\mathcal{U}\subseteq S, \left| \mathcal{U} \right|=s}({{G}_{\omega }}(\mathcal{U}){\mathbf{u}_{S}^{{\mathcal{U}}}})\oplus K_{S,\left| s \right|=s}^{(\beta)} {\mathbf{w}}$, in which ${{G}_{\omega }}({\mathcal{U}_{i}})=\varphi _{r\in {\mathcal{U}_{i}}}^{\omega }(W_{{{d}_{r,{\mathcal{U}_{i}}\backslash \{r\}}}}^{{{c}_{u}}})$, $\left| {\mathcal{U}_{i}} \right|=s$, and as a result of subpacketization of files, the size of file fragment $W_{{{d}_{r,{\mathcal{U}_{i}}\backslash \{r\}}}}^{{{c}_{u}}}$ is given by:
\begin{align}
\frac{F{{(q)}^{s-1}}{{(1-q)}^{K-s+1}}}{\left( \begin{smallmatrix}
	K-s \\ 
	\min (s+L-1,K)-s \\ 
	\end{smallmatrix} \right)}.
\end{align}

Thus, for a fixed value of $s$, the size of each coded message is given by:
\begin{align}
\underset{r\in \mathcal{U}_{i}}{\mathop{\max }}\,\left| W_{{{d}_{r,{\mathcal{U}_{i}}\backslash \{r\}}}}^{{{c}_{u}}} \right|=\frac{F{{(q)}^{s-1}}{{(1-q)}^{K-s+1}}}{\left( \begin{smallmatrix}
	K-s \\ 
	\min (s+L-1,K)-s \\ 
	\end{smallmatrix} \right)}.
\end{align}
As a result, each key $K_{S,\left| s \right|=s}^{(\beta)}$ must have the size of:
\begin{align}
&\left| K_{S,\left| s \right|=s}^{(\beta)} \right|=\underset{r\in \mathcal{U}_{i}}{\mathop{\max }}\,\left| W_{{{d}_{r,{\mathcal{U}_{i}}\backslash \{r\}}}}^{{{c}_{u}}} \right|=\frac{F{{(q)}^{s-1}}{{(1-q)}^{K-s+1}}}{\left( \begin{smallmatrix}
	K-s \\ 
	\min (s+L-1,K)-s \\ 
	\end{smallmatrix} \right)}.
\end{align}
where $\max$ operator implied that each sub-file is zero padded to the size of the largest sub-file in the set. Hence, the size of corresponding key is equal to the 
largest sub-file in the set.

For a fixed value of $s$, user $k$ requires the file fragments contained in $S\backslash \{k\}$ i.e., $\min (s+L-1,K)-1$ other users in the set $S$. We choose this set of $\min (s+L-1,K)-1$ users out of the rest of $K-1$ users. Since we need to repeat each coded message $\left(\begin{smallmatrix}
\min(s+L-1,K)-1 \\ 
s-1 \\ 
\end{smallmatrix}\right)$ times to derive different independent version of it to be decoded at the receiver side, we must generate $\left(\begin{smallmatrix}
K-1 \\ 
\min(s+L-1,K)-1 \\ 
\end{smallmatrix}\right)\left(\begin{smallmatrix}
\min(s+L-1,K)-1 \\ 
s-1 \\ 
\end{smallmatrix}\right)$ independent secret shared keys to have a unique key for each transmission. So, the total memory used by keys at each user's cache is calculated as:
\begin{align}
{{M}_{K}}=\sum\limits_{s=1}^{K}&{\frac{\left( \begin{smallmatrix}
		K-1 \\ 
		\min (s+L-1,K)-1 \\ 
		\end{smallmatrix} \right)\left( \begin{smallmatrix}
		\min (s+L-1,K)-1 \\ 
		s-1 \\ 
		\end{smallmatrix} \right)}{\left( \begin{smallmatrix}
		K-s \\ 
		\min (s+L-1,K)-s \\ 
		\end{smallmatrix} \right)}}\times{{(q)}^{s-1}}{{(1-q)}^{K-s+1}}, 
\end{align}
since finding a closed-form expression for ${M}_{K}$ is not straightforward, we find parameter $q$ in a numerical manner to meet the memory constraint.
\vspace{-4mm}
\subsection{Proof of Security}
In this subsection, we demonstrate how the delivery phase does not provide the eavesdropper with any information regarding the original files. The main idea underlying the proof of security is that the eavesdropper has not stored the keys to unlock the encrypted messages. In the delivery phase, legitimate clients can also quickly and error-free access the coded messages using their cached keys. We show that:
\begin{align}
I\left({y}_{e};{{W}_{1}},...,{{W}_{N}}\right)=0.
\end{align}
we have,
\vspace{-3mm}
\begin{align}
&I({y}_{e};{{W}_{1}},...,{{W}_{N}})=H({y}_{e})-H({y}_{e}|{{W}_{1}},...,{{W}_{N}})\\
&=H({y}_{e})-H(\{\mathbf{h}_{e}.{\mathbf{x}_{\omega }^s}\left(S\right):\left| S \right|=\min (s+L-1,K),\omega =1,...,\left( \begin{smallmatrix}
\min (s+L-1,K)-1 \\ 
s-1 \\ 
\end{smallmatrix} \right)\}_{s=1}^{K}|{{W}_{1}},...,{{W}_{N}})\label{eq-without-x-w-s}\\
&=H({y}_{e})-H(\{\mathbf{h}_{e}.\Bigl(\sum\nolimits_{\mathcal{U}\subseteq S, \left| \mathcal{U} \right|=s}{({{G}_{\omega }}(\mathcal{U}){\mathbf{u}_{S}^{{\mathcal{U}}}}})\oplus{K}_{S,\left| s \right|=s}^{(\beta)}{\mathbf{w}}\Bigl):\left| S \right|=\min (s+L-1,K),\nonumber\\
&\left| \mathcal{U} \right|=s,\omega =1,...,\left( \begin{smallmatrix}
\min (s+L-1,K)-1 \\ 
s-1 \\ 
\end{smallmatrix} \right)\}_{s=1}^{K}|{{W}_{1}},...,{{W}_{N}}) \label{eq-x-w-s}\\
&=H({y}_{e})-H(\{\mathbf{h}_{e}.\Bigl(\sum\nolimits_{\mathcal{U}\subseteq S, \left| \mathcal{U} \right|=s}{((\varphi _{r\in {\mathcal{U}_{i}}}^{\omega }(W_{{{d}_{r,{\mathcal{U}_{i}}\backslash \{r\}}}}^{{{c}_{u}}})){\mathbf{u}_{S}^{{\mathcal{U}}}}})\oplus {K}_{S,\left| s \right|=s}^{(\beta)}
{\mathbf{w}}\Bigl)\nonumber:\nonumber\\
&\left| S \right|=\min (s+L-1,K),\left| \mathcal{U} \right|=s,\omega =1,...,\left( \begin{smallmatrix}
\min (s+L-1,K)-1 \\ 
s-1 \\ 
\end{smallmatrix} \right)\}_{s=1}^{K}|{{W}_{1}},...,{{W}_{N}})\label{eq-G-w-s}\\
&=H({y}_{e})-H(\{{K}_{S,\left| s \right|=s}^{(\beta)}:\left| S \right|=\min (s+L-1,K),\omega =1,...,\left( \begin{smallmatrix}
\min (s+L-1,K)-1 \\ 
s-1 \\ 
\end{smallmatrix} \right)\}_{s=1}^{K}|{{W}_{1}},...,{{W}_{N}})\label{eq-k-w-s-decentralized}\\
&=H({y}_{e})-H(\{{K}_{S,\left| s \right|=s}^{(\beta)}:\left| S \right|=\min (s+L-1,K),\omega =1,...,\left( \begin{smallmatrix}
\min (s+L-1,K)-1 \\ 
s-1 \\ 
\end{smallmatrix} \right)\}_{s=1}^{K}),\label{eq-substitution-last}
\end{align}
where equation \eqref{eq-x-w-s} comes from replacing the coded message (${\mathbf{x}_{\omega }^s}$) in equation \eqref{eq-without-x-w-s}. One can reach equation \eqref{eq-G-w-s} by putting ${{G}_{\omega }}(\mathcal{U})$ in \eqref{eq-x-w-s}. By assuming that the statistics of the eavesdropper channel ($\mathbf{h}_{e}$) is known \cite{multi-eav,multi-eav-conference,Lightweight-channel}, and the $F$-bit discrete files are represented by $m$-bit symbols over finite field GF($q$), and considering the fact that beam-forming vectors of users (${\mathbf{u}_{S}^{\mathcal{U}}}$) are publicly-known variables, one can reach \eqref{eq-k-w-s-decentralized} from \eqref{eq-G-w-s}. As a result, the only term that offers randomness in the form of a key is, ${K}_{S,\left| s \right|=s}^{(\beta)}$ which takes up a part of the cache of each user as cost of security. In other words, keys ${K}_{S,\left| s \right|=s}^{(\beta)}$ provide the security of system. Furthermore, the last equality follows the fact that keys are uniformly distributed and are independent of the files $\left({{W}_{1}},...,{{W}_{N}}\right)$. Using the fact that
$H\left(A,B\right)\le H\left(A\right)+H\left(B\right)$.
We have:
\vspace{-2mm}
\begin{align}
&H\left({y}_{e}\right)=H\left(\{\mathbf{h}_{e}.{\mathbf{x}_{\omega }^s}\left(S\right)\}_{s=1}^{K}\right)\le \sum\limits_{s=1}^{K}H(\mathbf{h}_{e}.{\mathbf{x}_{\omega }^s}(S))\nonumber\\
& =\sum\limits_{s=1}^{K} \sum\limits_{i=1}^{\left( \begin{smallmatrix} 
	K \\ 
	\min (s+L-1,K) 
	\end{smallmatrix} \right)\left( \begin{smallmatrix} 
	\min (s+L-1,K)-1 \\ 
	s-1 
	\end{smallmatrix} \right)}H(\mathbf{h}_{e}.\Bigl(\sum\nolimits_{\mathcal{U}\subseteq {S}_{i}, \left| \mathcal{U} \right|=s}((\varphi _{r\in {\mathcal{U}_{i}}}^{\omega }(W_{{{d}_{r,{\mathcal{U}_{i}}\backslash \{r\}}}}^{{{c}_{u}}})){\mathbf{u}_{{S}_{i}}^{{\mathcal{U}}}})\nonumber\\
&\oplus{K}_{S_i,\left| s \right|=s}^{(\beta)}
{\mathbf{w}}\Bigl):\left| {S}_{i} \right|=\min (s+L-1,K),\left| \mathcal{U} \right|=s,\omega =1,...,\left( \begin{smallmatrix}
\min (s+L-1,K)-1 \\ 
s-1 \\ 
\end{smallmatrix} \right))\\
&=\sum\limits_{s=1}^{K}{\sum\limits_{i=1}^{\left( \begin{smallmatrix} 
		K \\ 
		\min (s+L-1,K) 
		\end{smallmatrix} \right)\left( \begin{smallmatrix} 
		\min (s+L-1,K)-1 \\ 
		s-1 
		\end{smallmatrix} \right)}{lo{{g}_{2}}}}\Bigl(\frac{F{{(q)}^{s-1}}{{(1-q)}^{K-s+1}}}{\left( \begin{smallmatrix}
	K-s \\ 
	\min (s+L-1,K)-s \\ 
	\end{smallmatrix} \right)}\Bigl)\\
&=\sum\limits_{s=1}^{K}{ \begin{pmatrix}
	K \\ 
	\min (s+L-1,K) \\ 
	\end{pmatrix}\begin{pmatrix}
	\min (s+L-1,K)-1 \\ 
	s-1 \\ 
	\end{pmatrix}}\times{log}_{2}\Bigl(\frac{F{{(q)}^{s-1}}{{(1-q)}^{K-s+1}}}{\left( \begin{smallmatrix}
	K-s \\ 
	\min (s+L-1,K)-s \\ 
	\end{smallmatrix} \right)}
\Bigl).\label{eq-h-y-e}
\end{align}
On the other hand, we have:
\vspace{-2mm}
\begin{align}
&H(\{{K}_{S,\left| s \right|=s}^{(\beta)}:\left| {S} \right|=\min (s+L-1,K),\omega =1,...,\left( \begin{smallmatrix}
\min (s+L-1,K)-1 \\ 
s-1 \\ 
\end{smallmatrix} \right)\}_{s=1}^{K})\\
&=\sum\limits_{s=1}^{K}H({K}_{S,\left| s \right|=s}^{(\beta)}:\left| {S} \right|=\min (s+L-1,K),\omega =1,...,\left( \begin{smallmatrix}
\min (s+L-1,K)-1 \\ 
s-1 \\ 
\end{smallmatrix} \right))\\
&=\sum\limits_{s=1}^{K}{\sum\limits_{i=1}^{\left( \begin{smallmatrix} 
		K \\ 
		\min (s+L-1,K) 
		\end{smallmatrix} \right)\left( \begin{smallmatrix} 
		\min (s+L-1,K)-1 \\ 
		s-1 
		\end{smallmatrix} \right)}}H({K}_{S_i,\left| s \right|=s}^{(\beta)}:\left| {S}_{i} \right|=\min (s+L-1,K)\nonumber\\
	&,\omega =1,...,\left( \begin{smallmatrix}
\min (s+L-1,K)-1 \\ 
s-1 \\ 
\end{smallmatrix} \right))\\
&=\sum\limits_{s=1}^{K}{\sum\limits_{i=1}^{\left( \begin{smallmatrix} 
		K \\ 
		\min (s+L-1,K) 
		\end{smallmatrix} \right)\left( \begin{smallmatrix} 
		\min (s+L-1,K)-1 \\ 
		s-1 
		\end{smallmatrix} \right)}{lo{{g}_{2}}}}\Bigl(\frac{F{{(q)}^{s-1}}{{(1-q)}^{K-s+1}}}{\left( \begin{smallmatrix}
	K-s \\ 
	\min (s+L-1,K)-s \\ 
	\end{smallmatrix} \right)}\Bigl)\\
&=\sum\limits_{s=1}^{K}{ \begin{pmatrix}
	K \\ 
	\min (s+L-1,K) \\ 
	\end{pmatrix}\begin{pmatrix}
	\min (s+L-1,K)-1 \\ 
	s-1 \\ 
	\end{pmatrix}} \times {log}_{2}\Bigl(\frac{F{{(q)}^{s-1}}{{(1-q)}^{K-s+1}}}{\left( \begin{smallmatrix}
	K-s \\ 
	\min (s+L-1,K)-s \\ 
	\end{smallmatrix} \right)}
\Bigl),\label{eq-last}
\end{align}
where the last equality in \eqref{eq-last} follows from the fact that the keys are orthogonal to each other and they are uniformly distributed. Substituting \eqref{eq-h-y-e}, \eqref{eq-last} into \eqref{eq-substitution-last}, and considering the fact that for any $X$ and $Y$, $I\left({X};{Y}\right) \ge 0$, we have:
\vspace{-2mm}
\begin{align}
I\left({y}_{e};{{W}_{1}},...,{{W}_{N}}\right)=0.
\end{align}	
% Can use something like this to put references on a page
% by themselves when using endfloat and the captionsoff option.
\ifCLASSOPTIONcaptionsoff
  \newpage
\fi

% biography section
% 
% If you have an EPS/PDF photo (graphicx package needed) extra braces are
% needed around the contents of the optional argument to biography to prevent
% the LaTeX parser from getting confused when it sees the complicated
% \includegraphics command within an optional argument. (You could create
% your own custom macro containing the \includegraphics command to make things
% simpler here.)
%\begin{IEEEbiography}[{\includegraphics[width=1in,height=1.25in,clip,keepaspectratio]{mshell}}]{Michael Shell}
% or if you just want to reserve a space for a photo:

%\begin{IEEEbiography}{Michael Shell}
%Biography text here.
%\end{IEEEbiography}

% if you will not have a photo at all:
%\begin{IEEEbiographynophoto}{John Doe}
%Biography text here.
%\end{IEEEbiographynophoto}

% insert where needed to balance the two columns on the last page with
% biographies
%\newpage

%\begin{IEEEbiographynophoto}{Jane Doe}
%Biography text here.
%\end{IEEEbiographynophoto}

% You can push biographies down or up by placing
% a \vfill before or after them. The appropriate
% use of \vfill depends on what kind of text is
% on the last page and whether or not the columns
% are being equalized.

%\vfill

% Can be used to pull up biographies so that the bottom of the last one
% is flush with the other column.
%\enlargethispage{-5in}

% that's all folks
\end{document}